%% AVERAGING FOR MAPS + BEAM DYNAMICS
%% 08/01/03 Reformatted for DESY red report
%% 11/04/03 Reormatted to include figures from APS 02 slides
\input epsf

\hsize=17cm
\hoffset=-.5cm
\vsize=24.2cm
\voffset=0cm

\def\D{{\cal D}}
\def\eps{\varepsilon}
\def\fbar{\overline{f}}
\def\fhat{\widehat{f}}
\def\fwig{\widetilde{f}}

\def\Hchat{\widehat{\cal H}}
\def\K{{\cal K}}
\def\R{{\bf R}}
\def\N{{\bf N}}

\def\Rn{{{\bf R}^d}}

\def\To{{\bf T}^1}

\def\yhat{\check{y}}
\def\zhat{\check{z}}
\def\Z{{\bf Z}}

\def\transp{^{\rm T}}
\def\CT{\zeta}

\def\noi{\noindent}
\def\disp{\displaystyle}
\def\negskip{\vskip -.02 true in}  %%  (also to be modified)
\def\vn{\vskip -.03 true in\noindent}

\font\tw = cmtt9

\centerline{\bf First-Order Averaging Principles for Maps}

\smallskip
\centerline{\bf with Applications to Beam Dynamics in Particle
Accelerators}

\bigskip\medskip

{\settabs\+\hskip .1 true in & type in something about
this long & type in something about this long & \cr

\+&H.\ Scott Dumas&James A. Ellison &Mathias Vogt\cr

\vskip .02 true in
\+&{\sl Department of Mathematics}&{\sl Department of Mathematics}&
{\sl Deutsches Elektronen Synchrotron}\cr

\negskip
\+&{\sl University of Cincinnati}&{\sl University of New Mexico}&
{\sl Notkestra\ss e 85}\cr

\negskip
\+&{\sl Cincinnati, OH 45221--0025}&{\sl Albuquerque, NM 87131}&
{\sl 22607 Hamburg, Germany}\cr

\+&{\tw scott.dumas@uc.edu}&{\tw ellison@math.unm.edu}&
{\tw vogtm@mail.desy.de}\cr

}

\bigskip\medskip
\noi{\bf Abstract.}  {\sl
For slowly evolving, discrete-time-dependent systems of difference
equations (iterated maps),
we believe the simplest means of demonstrating the validity
of the averaging method at first order is by way of a lemma that we
call Besjes' inequality.  In this paper, we develop the Besjes
inequality for identity maps with perturbations that are (i) at
low-order resonance (periodic with short period)  and (ii) far from
low-order resonance in the discrete time.  We use these inequalities
to prove corresponding first-order averaging principles, together
with a principle of adiabatic invariance on extended timescales; and we
generalize and apply these mathematical results to model problems in
accelerator beam dynamics, and to the H\'enon map.}

\medskip
\noi{\bf Keywords:} Averaging method; averaging principle; difference
equations; iterated maps; small divisors; adiabatic invariance; accelerator
beam dynamics; kick-rotate model; H\'enon map

\medskip
\noi{\bf 2000 MSC:} 39A11; 37J40; 70K65; 70H11; 78A35; 70Fxx

\medskip
\noi{\bf Submitted to}\ SIAM Journal of Applied Dynamical Systems

%[stablty \&\ asymptcs of diffnc eqs/ pertbns, nrml frms, smll dvsrs/
%avgng of prtbns/ adiabtc invrnts/ motion of chrgd prtcls/ dynam of
%particle systms]

\bigskip\medskip\noi
{\bf 1. Introduction}

\medskip\noi
In broadest terms, the method of averaging (or ``averaging
principle") may be described as follows: to approximate the evolution
of a system with motions occurring on both fast and slow timescales,
one uses a simpler system obtained by somehow averaging over the fast
motion of the original system.  In the context of difference equations
(or ``iterated maps"), the most elementary situation to which the method
applies occurs in periodic systems of the form
$$
x_{n+1}
\,=\,
x_n \, + \,
\eps f(x_n,n)
\eqno(1.1)
$$\vn
where $x_n\in U\subset\Rn$, $n\in\N$, $\eps>0$ is a small parameter,
and $f:U\times\N\to\Rn$ is a bounded, locally $x$-Lipschitz,
discrete-time-dependent function of period $p$ in $n$.  Solutions of
system (1.1) are approximated by solutions of the associated averaged
system
$$
y_{n+1}
\,=\,
y_n \, + \,
\eps \fhat(y_n)
\eqno(1.2)
$$\vn
where the autonomous function $\fhat:U\to\Rn\,$ \
(the {\sl average of} $f$) is given by
$\,\fhat(y) = (1/p)\sum_{n=0}^{p-1} f(y,n)$. In this context the
averaging principle asserts that solutions $x_n$ of Eq.\ (1.1) and $y_n$
of Eq.\ (1.2) that start at the same initial condition remain
$O(\eps)$-close on a discrete timescale of $O(1/\eps)$.
It is also often useful to use the continuous-time solutions of the
corresponding averaged ODE
$$
 {dy\over dt}    \,=\, \eps \fhat(y)
% \qquad\quad{\rm (or\ equivalently}\quad
% {d\yhat\over dt'}\,=\, \fhat(\yhat)\quad
% {\rm with}\quad t'=\eps t {\rm )}
 \eqno(1.3)
$$\vn
to approximate the discrete-time solutions of Eq.\ (1.2) and hence also
those of Eq.\ (1.1), so that we obtain the two approximation relations
$x_n = y_n+O(\eps)$ and $x_n = y(n)+O(\eps)$ for $0\le n\le O(1/\eps)$
(note that $y_n$ and $y(n)$ have different meanings).  A more precise
formulation appears below in Theorem 1, followed by a very elementary
proof that makes no use of the usual transformation that appears in
textbooks (it is not always recognized that first-order averaging may
be justified without the sort of coordinate transformations used, for
example, in canonical perturbation theory).

Equation (1.1) is a special case of a more general problem on which we
focus in this paper.  Let $\nu\in\R$, $U\subset\Rn$, and
$f:\,U\times\R\to\Rn$ be periodic with period 1 in its second argument.
We then consider the system
$$
 x_{n+1} = x_n +\eps f(x_n,n\nu)\,. \eqno(1.4)
$$\vn
The analysis of this problem is similar to the analysis of the
flow problem $dx/dt=\eps f(x,t)$ when $f$ is quasiperiodic in $t$
with two base frequencies, since small divisors enter both problems in the
same way.  Clearly Eq.\ (1.4) reduces to Eq.\ (1.1) when $\nu=q/p$ is
rational.
For $\nu$ irrational, we know from Weyl's equidistribution theorem [K\"o]
that the average of $f(x,n\nu)$ over $n$ exists and equals
 $\fbar(x)=\int_0^1 f(x,t)\,dt$.
It is therefore natural to ask for what values of $\nu$ the solutions of
 Eq.\ (1.4) can be approximated by solutions of the two systems
$$
 y_{n+1} = y_n + \eps \fbar(y_n) \eqno(1.5)
$$\vn
 and
$$
 {dy\over dt} \,=\, \eps \fbar(y) \, .
% \qquad\quad{\rm (or\ equivalently}\quad
% {d\yhat\over dt'}\,=\, \fbar(\yhat)\quad
% {\rm with}\quad
% t'=\eps t {\rm )}
 \eqno(1.6)
$$\vn
In answering this question, it also seems natural (from the mathematical
viewpoint) to introduce Diophantine conditions on $\nu$, but these
conditions in their usual form are problematic in applications, and not
wholly necessary, as we shall see.  In fact, we present approximation
theorems that are both theoretically satisfying and suited to
applications.  In particular, we weaken the usual small divisor
conditions on $\nu$ (in which $\nu$ satisfies infinitely many
``Diophantine conditions''), requiring instead only finitely many
conditions at appropriately low order.  These conditions exclude $\nu$
from zones centered on low-order rationals, and in this
``far-from-low-order-resonance case'' (where $\nu$ satisfies only
``truncated Diophantine conditions'' and is {\it not\/} necessarily
irrational), we again find that
$x_n=y_n+O(\eps)=y(n)+O(\eps)$ for $0\le n\le O(1/\eps)$ (see Theorem 2
below). Under the additional hypothesis that the average of the
perturbation vanishes, we are able to show adiabatic invariance of
solutions of system (1.4) on extended timescales up to $O(1/\eps^2)$ (see
Theorem 3).  We thus have results for both low-order resonant (or rational)
$\nu$, and for $\nu$ far from low-order resonance.

Finally, a simple trick permits us to explore $O(\eps)$ neighborhoods of
low-order resonances $\nu=q/p$: we set $\nu=q/p+\eps a$ (where $a\in\R$
should be viewed as a measure of the $O(\eps)$ displacement from the
resonance) and rewrite Eq.\ (1.4) as the system
$$ \pmatrix{ x_{n+1} \cr \tau_{n+1} }  =
   \pmatrix{
   x_n + \eps \, f(x_n, {q\over p}n+\tau_n) \cr
   \tau_n + \eps a}\,.  \eqno(1.7)
$$\vn
This is in the form of Eq.\ (1.1) with $x_n$ replaced by
$(x_n,\tau_n)\transp$.
Writing $\fhat(x,\tau)=1/p\,\sum_{n=0}^{p-1}f(x,nq/p+\tau)$,
the averaged problem reduces to
$$
 \pmatrix{ y_{n+1} \cr \tau_{n+1} }  =
 \pmatrix{ y_n + \eps\, \fhat(y_n,\tau_n)  \cr
           \tau_n + \eps a                    } \,,
\eqno(1.8)
$$\vn
and we recapture the relations $x_n=y_n+O(\eps) = y(n) + O(\eps)$ \
for \ $0\le n\le O(1/\eps)$, where $y(t)$ is the solution of the system
$$
 {d\over dt}\,\pmatrix{ y \cr \tau }  = \eps
        \pmatrix{\fhat(y,\tau) \cr a } \,, \eqno(1.9)
$$\vn
which is equivalent to the
non-autonomous system \ $dy/dt=\eps\fhat(y,\eps a t)\,$; see Proposition C
below. 

\headline{\centerline{Averaging for Maps}}

Initially, we state Theorems 1, 2, and 3 under the hypothesis that the
perturbation $\eps f$ has compact support in its $x$-domain, which is
assumed to be all of $\Rn$; this avoids {\sl a priori\/} restrictions on
$\eps$ and permits clear proofs.  To obtain results better suited to
applications, we then give propositions that extend our theorems to more
general perturbations on more general domains, and also to more general
Diophantine conditions in which the zones mentioned above are allowed to
depend on $\eps$; this in turn allows $\nu$ to come within $O(\eps^\lambda)$
of low-order rationals, but with loss of accuracy in the approximation
(see Propositions A and B below).
Using the generalized versions of our theorems (provided by Propositions
A, B, and C), we obtain an essentially complete description of solutions
of system (1.4) on $O(1/\eps)$ timescales for various values of $\nu$
(there are however thin gaps at the boundaries between the $\nu$ for
which resonant and nonresonant motions occur; cf.\ Remark 2.5 below).

From the viewpoint of applied mathematics, perhaps the most interesting
aspect of our results is that our Theorems 2 and 3 have physically
realistic, truncated Diophantine conditions {\it in their hypotheses},
yet provide approximations {\it valid on full $O(1/\eps)$ time intervals}.
For more general multiphase averaging principles, such nice hypotheses
lead to passage through resonance, and thus to approximations that are
valid only on somewhat shorter time intervals (cf.\ [ABG]); but we have
identified an important class of simpler problems arising from accelerator
beam dynamics in which both the realistic hypotheses and the full
$O(1/\eps)$ validity times can coexist.

More generally, averaging principles for maps are not new; results in this
direction have been available since the 1960s (cf.\ for example [Bel],
[Dr]).  However, a detailed theory of Eq.\ (1.4) suitable for applications
appears to be missing from the literature, and we proceed to fill that gap
in this paper.  We do not however illustrate the full range of
applicability of our theorems; instead we discuss a single important
example from the class of problems which motivated this investigation,
namely the so-called ``kick-rotate'' models from accelerator dynamics,
represented by
$$
 w_{n+1} = M (w_n + \eps K(w_n)),
$$
which takes the form of Eq.\ (1.4) under the transformation
$w_n = M^n x_n$.
In this paper, we emphasize this model's application to the so-called
weak-strong beam-beam interaction (see \S 3.2 below),
but kick-rotate models also apply to other localized perturbations
in accelerators.

We point out that our discussion below in Section 3 is the first
 mathematically rigorous treatment of this important class of models in the
 sense of asymptotics.
Many beam dynamics treatments start with a smooth Hamiltonian formulation and
 apply canonical perturbation theory without rigorous error analysis.
Resonances are often not treated in the spirit of
 perturbation theory (see however the paper [Ru] for
 a nice discussion of the use of perturbation theory in beam dynamics).
Futhermore, delta function perturbations are often used in this smooth
 Hamiltonian framework (it is of course more natural to use them with maps),
 making the validity of any resulting approximations hard to assess.
(The paper [CBW] gives a nice introduction to the beam-beam interaction,
 but uses this Hamiltonian/delta function approach.)
One notable exception to the Hamiltonian formulation is the work on maps
 using Lie operators, a good
 discussion of which may be found in [Fo], where the author
 has carried this approach quite far---to realistic machine models---but
 without focusing on rigorous asymptotics.
We are aware of another research group working on highly mathematical
 perturbation treatments of beam dynamics in the context of maps [BGSTT]
 but our work here is quite distinct from theirs.
To begin with, our perturbation parameter is the size of the ``kick''
 (cf.\ Section 3.1 below), whereas they study the long
 time stability of the origin (which is assumed to be a linearly stable
 elliptic fixed point), using the distance from the origin as a
 perturbation parameter.
Futhermore, their analysis is quite complex, as they pursue Nekhoroshev-type
 results involving many successive coordinate transformations
 which give rise to complicated and restrictive hypotheses that may be
 difficult to verify in practice.
In our own approach, resonances are treated in the simplest possible rigorous
 way, and we obtain a natural partition of ``tune space'' into regions
 with distinct resonance properties.  We believe this is an important new
 feature, both conceptually and practically.
Of course, it is important to note that our method gives approximations
 to leading order only (using no transformations, as mentioned earlier);
 this accounts for much of its radical simplicity.
It also allows us to use simple and realistic hypotheses, in turn permitting
 meaningful comparison of the kick-rotate approximation with numerical
 experiments.
Overall, we believe that our treatment provides the starting point for a
 simple, effective means of studying mathematical models of beam dynamics
 rigorously, and that its development should complement
 previous theoretical and mathematical work.

The remainder of this paper is organized as follows.  In Section 2
we present the details of our averaging results described informally
above.  In Section 3 we apply the averaging principles to model problems
in accelerator beam dynamics, showing that solutions of a class of
``kick-rotate" models are well-approximated by solutions of the
corresponding averaged models. We also apply the adiabatic invariance
principle to the H\'enon map (often used to model sextupole magnets in
accelerators). In Section 4, we formulate the main technical tools
required to prove the results in Section 2.  These are the so-called
Besjes inequality for periodic functions (Lemma 1, \S 4.1), and  its
generalization to functions far from low-order resonance (Lemma 2,
\S 4.2.2).  After formulating and proving these inequalities,
we use them to prove the mathematical results from Section 2.
Finally, for the sake of completeness, in the Appendix we state and
prove two elementary results used in earlier proofs.

We end this introduction with a few words about notation.  We use the
symbols $\N$, $\R$, $\R_+$, and $\Z$ to denote, respectively, the
counting numbers $\{0,1,2,\ldots\}$, the real numbers, the positive
real numbers, and the integers.  The symbol $|\ |$ indicates the
Euclidean norm on $\Rn$ (or the absolute value $|k|$ of an integer
$k$), and $\|\ \|_S$ denotes the uniform norm of a function over the
set $S$; i.e., $\|F\|_S:=\sup_{x\in S}|F(x)|$.

%%% \vfill\eject %%% pagebreak

\bigskip\medskip\noi
{\bf 2. \ Averaging Principles and Adiabatic Invariance}

\medskip\noi
In this section we state---and provide brief remarks on---our
approximation results for maps as discussed in the introduction above.

\bigskip\noi
{\bf 2.1 Averaging for Maps with Periodic Perturbations}

\smallskip
Let us be more precise about the functions $f$ in Eq.\ (1.1)
to which our results apply.  First, taking $S=\Rn\times\N$,
we assume that $f:S\to\R$ satisfies the following:

\smallskip
(i) $f$ is bounded on $S$ and $f(\cdot,n)$ is locally Lipschitz,
uniformly in $n$

(ii) There exists a positive integer $p$ such that \ $(x,n)\in
S\Rightarrow f(x,n+p) = f(x,n)$

(iii) There is an $r>0$ such that $|x|\ge r$
and $n\in\N\Rightarrow f(x,n)=0$

\medskip
\noi When $f$ satisfies (ii), we say it is ``periodic with
period $p$ in its second argument"; and when it satisfies (iii), it is
``compactly supported in $x$, uniformly in $n$."
It follows from (i) and (iii) that $f$ is globally Lipschitz in $x$, uniformly
in $n$.  In Subsection 2.4 we
show how to treat the case where $f$ is not compactly supported.

We now state a simple averaging principle for maps with periodic
perturbation $\eps f(x,n)$ and corresponding averaged perturbation
$\,\eps\fhat(y)=(\eps/p)\sum_{n=0}^{p-1}f(y,n)\;$:

\proclaim Theorem 1.
Let $S=\Rn\times\N$, and suppose $f:S\to\Rn$ satisfies assumptions
 (i), (ii), and (iii) above.
Fix $\eps\in(0,1]$, and consider the system
$$
  x_{n+1} \,=\, x_n \, + \, \eps f(x_n,n)
\eqno(1.1)
$$
 together with the associated averaged systems
$$
  y_{n+1} \,=\, y_n \, + \, \eps \fhat(y_n) \ \qquad (1.2)\,,
  \qquad\qquad\quad {\rm and} \qquad\qquad\quad
 {dy\over dt} \,=\, \eps \fhat(y) \,.\ \qquad (1.3)
$$
Choose $T>0$ to capture the desired properties of system (1.3) on
$[0,T/\eps]$. Then there exist positive constants $C=C(T)$ and $C'=C'(T)$
  %(independent of the period $p$)
such that the solutions $x_n$, $y_n$,
and $y(t)$ of Eqs.\ (1.1), (1.2), and (1.3) with common initial condition
$x_0=y_0=y(0)$ exist uniquely for all time and satisfy $|x_n-y_n|\le
Cp\,\eps$ \ and $|x_n-y(n)|\le (Cp+C')\,\eps$ \ for  $0\le n\le T/\eps$.

\bigskip\noi
{\bf 2.2 Averaging for Maps With
Perturbations Far From Low-Order Resonance}

\smallskip
We now present an averaging principle for system (1.4), where $\nu$ is a
fixed positive number.
When we write $\nu=q/p$, we mean that $q$ and $p>0$ are relatively prime integers
 with the {\sl order\/} of the rational number $\nu$ given by $p>0$.
Using this convention, we first note that if $\nu=q/p$,
 then $f(x,n\nu)$ has integer period $p$ in $n$, and
Theorem 1 applies. In fact, as we shall see in Proposition C, Theorem 1
applies not only {\it at\/} low-order rationals but also {\it near\/}
them. However, since the error estimate in this theorem is proportional
to $p$, it is not very useful when $p$ is ``large.''
We therefore restrict use of Theorem 1 to situations where $p$ is
``small'' (the ``low-order-resonance case"), and we next focus on
situations where $\nu$ is far from low-order rational
numbers (the ``far-from-low-order-resonance case").
In this case small divisors inevitably enter the analysis
(see the proof of Lemma 2, \S4.2.2) and it might be
expected that $\nu$ would need to be ``highly irrational"
(e.g.\ satisfy infinitely many Diophantine conditions).
We show instead that the averaging principle may be
established when $\nu$ satisfies only finitely many Diophantine
conditions to a certain order, and we call these {\sl truncated
Diophantine conditions}.

In more precise terms, $\nu$ satisfies truncated Diophantine
conditions if it belongs to the set $\D(\phi,R)$ defined below in
Eq.\ (4.3), where $\phi$ is the {\sl zone function\/} of the
Diophantine condition and $R>0$ is the {\sl truncation order\/} or
{\sl ultraviolet cutoff}, which gives precise meaning to the
phrase ``$p$ large'' used above (i.e., $p$ is large if $p>R$).
Roughly speaking, $\D(\phi,R)$ is constructed by
removing open intervals centered on low-order rationals $\nu=q/p$.
The zone function $\phi$ controls the size of the intervals removed, and
the cutoff $R$ is the maximal order of rationals from around which
intervals are removed.
These terms are defined precisely in Subsection 4.2.1
(to fully understand
the difference between truncated and ordinary
Diophantine conditions, and to appreciate the advantages offered by
the former, the reader may find it worthwhile to read that subsection).

With truncated Diophantine conditions given explicitly in Eq.\ (4.3),
we now consider the
class of functions to which our next result applies.  For
$S=\Rn\times\R$ we consider functions
$f:S\to\Rn$ satisfying the following conditions (analogous to (i)
through (iii) in \S 2.1):

\smallskip
(j) $f$ is of class $C^4$ on $S$

(jj) $(x,\theta)\in S\Rightarrow f(x,\theta + 1) = f(x,\theta)$

(jjj) There is an $r>0$ such that $|x|\ge r$
and $\theta\in\R\Rightarrow f(x,\theta)=0$

\medskip
Terminology for describing conditions (jj) and (jjj) is similar to that
for describing conditions (ii) and (iii) above in Subsection 2.1.  Since
we assume $f$ has unit period in its second argument, its average $\fbar$
is simply $\fbar(y):=\int_0^1 f(y,\theta)\,d\theta$.  Finally, we alert
the reader that the truncated Diophantine conditions satisfied by $\nu$
must be adapted to $f$ in the sense that the zone function $\phi$ must
decay appropriately; this is made precise in Eq.\ (4.2) of Subsection
4.2.1 (basically $\phi$ must decay fast enough so that $\D(\phi,R)$ is
nonempty, but slow enough so that the series in Eq.\ (4.2) converges; this
accounts for assumption (j) above and our specific choice of $\phi$ as
discussed in \S4.2.1).

\vfill\eject %%% pagebreak

We now state our averaging principle for maps with perturbations
$\,\eps f(x,n\nu)\,$ far from low-order resonance and averaged perturbation
$\eps\fbar(y)$ as above:

\proclaim Theorem 2.  Let $S=\Rn\times\R$, suppose $f:S\to\Rn$ satisfies
assumptions (j), (jj), and (jjj) above, and suppose the zone function
$\phi$ is adapted to $f$ on $\Rn$ in the sense of Eq.\ (4.2).  Fix
$\eps\in(0,1]$, and consider the system
$$
 x_{n+1} \ = \ x_n + \eps f(x_n,n\nu)
\eqno(1.4)
$$
together with the associated averaged systems
$$
y_{n+1} \, = \, y_n + \eps \fbar(y_n)
\qquad (1.5)\,,
\qquad\qquad\quad {\rm and} \qquad\qquad\quad
{dy\over dt} \,=\, \eps\fbar(y)\,.
 \qquad (1.6)
$$
Choose $T>0$ to capture the desired properties of system (1.6) on
$[0,T/\eps]$.  Then there exist positive constants $R_\eps$, $C=C(f,\phi,T)$,
and $C'=C'(f,\phi,T)$ such that whenever $\nu\in\D(\phi,R_\eps)$
(defined in Eq.\ (4.3)), the solutions $x_n$, $y_n$, and $y(t)$ of Eqs.\
(1.4), (1.5), and (1.6) with common initial condition $x_0 = y_0 = y(0)$
exist uniquely for all time and satisfy $|x_n-y_n|\le C\,\eps$ \ and \
$|x_n-y(n)|\le C'\,\eps$ \ for \ $0\le n\le T/\eps$.

\smallskip
\noi{\bf Remark 2.1} \ For averaging principles of this type, it is
natural to consider the average \hfill\break
 $\lim_{N\to\infty}(1/N)\sum_{n=0}^{N-1} f(x,n\nu)$ \ of $f$ over $n$
 as mentioned in the introduction. Under mild integrability conditions on
$f$, it can be shown that when $\nu$ is irrational, this average converges
to $\int_0^1 f(x,\theta)\,d\theta$, which is the average used here
 (this is related to Weyl's equidistribution theorem; cf.\ [Br] and
[K\"o]).  However, our results do not require the existence of the average
of $f(x,n\nu)$ over $n$, nor do they require $\nu$ to be irrational;
 instead we require $\nu\in\D(\phi,R_\eps)$, and this latter set contains
 many rationals of order greater than $R_\eps$.

\bigskip\noi
{\bf 2.3 Adiabatic Invariance on Extended Timescales}

\smallskip
In this subsection, we consider a special system somewhat like a
perturbation of an integrable Hamiltonian system.  As in Theorem 2,
we assume that $\nu$
satisfies truncated Diophantine conditions, but now we assume
additionally that the perturbation $\eps f$  has zero mean; i.e.,
we assume that

\smallskip
(jw) For each $x\in\Rn$, \ $\disp\int_0^1 f(x,\theta)\,d\theta = 0$

\medskip
This extra hypothesis gives an averaging principle showing that
the action-like variables are adiabatically invariant over timescales
longer than $O(1/\eps)$:

\proclaim Theorem 3.  Let $S=\Rn\times\R$, suppose $f:S\to\Rn$
satisfies conditions (j), (jj), (jjj), and (jw) above, and suppose
the zone function $\phi$ is adapted to $f$ on $\Rn$ (as in
Eq.\ (4.2)).  Fix $\eps\in(0,1]$, choose $T>0$, and consider the system
$$
 x_{n+1} \, = \, x_n + \eps f(x_n,n\nu) \eqno(1.4)
$$
with arbitrary initial condition $x_0\in\Rn$.  Then there exist positive
constants $R_\eps$, $K_1=K_1(f,\phi)$, and $K_2=K_2(f,\phi)$ such that
whenever $\nu\in\D(\phi,R_\eps)$ (cf.\ Eq.\ (4.3)), the solution $x_n$ of
Eq.\ (1.4) satisfies $|x_n - x_0| \ \le \ K_1\,\eps + K_2\,\eps^2 n$ for
$n\in\N$.  In particular, for $0\le\alpha\le1$, we have
$|x_n - x_0| \le C(T)\,\eps^\alpha$ for $0< n\le T/\eps^{2-\alpha}$,
where $C(T)=K_1+K_2T$.

\smallskip
\noi{\bf Remark 2.2} \ Using second (or higher) order averaging, it is
possible to get a better estimate of $|x_n - x_0|$ on the full
$O(1/\eps^2)$ time interval (see [ES] for a flow version).

\bigskip\noi
{\bf 2.4 Extensions and Generalizations}

\smallskip
In this subsection we give three propositions that extend and generalize
our results above, making them more suitable for applications.
Our first proposition shows that Theorems 2 and 3 may be generalized
to the case where the zones of the truncated Diophantine conditions
depend on $\eps$.

\proclaim Proposition A ($\eps$-dependent zone functions).  Suppose that
$0\le\lambda\le1$, and that in Theorem 2 [or Theorem 3], the zone function
$\phi$ is replaced by the new zone function $\eps^\lambda\phi$.  Then the
conclusions of the theorem remain true, provided that the error estimates
$C\,\eps$ and $C'\eps$ are modified to read $C\eps^{1-\lambda}$ and
$C'\eps^{1-\lambda}$ [or $C(T)\,\eps^\alpha$ is modified to read
$C(T)\,\eps^{\alpha-\lambda}$].

In order to clarify and simplify the mathematical structure of our
methods, we have presented Theorems 1, 2, and 3 under the assumption
that the perturbations have compact support on spatial domains that
are all of $\Rn$.  Our next proposition shows that this assumption may be
removed at little cost.

\proclaim Proposition B (more general perturbations).
Suppose that the domain $S=\Rn\times\N$ in Theorem 1 is replaced
by the more general domain $S'=U\times\N$, where $U\subset\Rn$ is open
[or the domain $S=\Rn\times\R$ in Theorem 2 or 3 is replaced
by $S'=U\times\R$, $U\subset\Rn$ open], and assumption (iii) is
removed from the hypotheses of Theorem 1 [or (jjj) is removed from
the hypotheses of Theorem 2 or 3].  Then the conclusions of
Theorem 1 [or Theorem 2 or 3] remain true provided that: \
(a) $0\le \eps<\eps_0$, where the threshold $\eps_0>0$
may be estimated as outlined below in Subsection 4.3.2; and \
(b) the conclusion ``exist uniquely for all time" is replaced by ``exist
uniquely on the time interval $[0,T/\eps]$," with $T>0$ chosen strictly
less than $\beta(x_0)$, where $[0,\beta(x_0))$ is the maximal forward
interval of existence for the averaged flow problem
$d\yhat/dt'=\fhat(\yhat)$ in the domain $U$ [or for the flow problem
$d\yhat/dt'=\fbar(\yhat)$ in $U$].

\smallskip
\noi{\bf Remark 2.3} \ Of course Proposition A also applies to Proposition B.

\medskip
The following proposition shows that Theorem 1 may be used to analyze
the dynamics of solutions of Eq.\ (1.4) in $O(\eps)$ neighborhoods of
low-order resonances $\nu=q/p$.

\proclaim Proposition C (behavior near low-order resonance).
Let $U\subset\Rn$ be open, $S'=U\times\R$, and suppose $f:$ $S'\to\Rn$
satisfies conditions (j) and (jj) of Theorem 2 with $S$ replaced by $S'$.
Fix the rational number $q/p$, $p>0$ and $q$ relatively prime, and fix
$a\in\R$. Then Eq.\ (1.4) with $\nu=q/p+a\,\eps$ may be rewritten as Eq.\
(1.7), and Theorem 1 together with Proposition B apply with $x$ and $y$
replaced by $(x,\tau)\transp$ and $(y,\tau)\transp$ respectively.  In
particular there are positive constants $\eps_0$, \ $c=c(T,|a|)$, and
$c'=c'(T,|a|)$ such that $|x_n-y_n|\le c\,p\,\eps$ and
$|x_n-y(n)|\le (cp+c')\,\eps$ \ for \ $0\le\eps<\eps_0$ \ and \
$0\le n\le T/\eps$.

\smallskip
\noi{\bf Remark 2.4} \ Clearly $y_n$ evolves by
 $y_{n+1}=y_n+\eps\fhat(y_n,\eps a n)$; and also $y(n)=\yhat(\eps n)$,
 where $\yhat$ evolves via $d\yhat/dt=\fhat(\yhat,a t)$.

\medskip
\noi{\bf Remark 2.5} \ Propositions A and B characterize the motion of $x_n$
to within $O(\eps^{1-\lambda})$ for $\nu$ away from low-order rationals,
i.e., outside of $O(\eps^\lambda\phi(p)/p)$ neighborhoods of rationals $q/p$
with $0<p\le R_\eps$.  For these $\nu$ the nonresonant normal form of Eq.\
(1.6) applies. Proposition C characterizes the motion to within $O(\eps p)$ for
$\nu$ inside $O(\eps)$ neighborhoods of $q/p$.  For these $\nu$ the resonant
normal form of Eq.\ (1.9) applies. What is missing is information about the
motion for $\nu$ in the gaps between the domains of validity of the resonant
normal form and the nonresonant normal form.  The size of the gaps decreases
to zero as $\lambda\nearrow1$; however, the error in the nonresonant normal
form simultaneously deteriorates to $O(1)$.  High-order rationals, i.e.\
$q/p$ with $p>R_\eps$, are of course treated using Proposition B.  It is
interesting to note that they may also be treated using Proposition C;
however, the $O(p\eps+\eps)$ error bound deteriorates to $O(1)$ as $p$
approaches $O(1/\eps)$.

%%%\vfill\eject  %%% pagebreak

\bigskip\medskip\noi
{\bf 3. \ Examples from Accelerator Beam Dynamics}

\medskip\noi
Modern particle accelerators operate at the limits of current technology,
and their design and operation depend crucially on an understanding of
the dynamics of particle beams.
% A number of mathematical models of beam
% dynamics are used, and highly sophisticated numerical and analytical
% tools are in turn brought to bear on these models.
In this section we
give examples showing how Theorems 1 and 2 (supplemented by Propositions
A, B and C) may be used to analyze a class of beam dynamics models, and
how Theorem 3 may be used to analyze the H\'enon map (which is itself a
model of certain features in beam dynamics).  In fact, our averaging
principles for maps have features that make them especially effective for
this purpose; namely, they compare solutions of the exact and averaged
model problems in the simplest possible way, and produce rigorous
mathematical bounds on the difference between these solutions in an
essentially optimal fashion.  Although $O(1/\eps)$ times may be short by
accelerator standards (and adiabatic invariance of actions on
$O(1/\eps^2)$ times is perhaps ideal), we see our work here as an
important step in understanding the dynamics of maps on long timescales.
We emphasize that these are rigorous error bounds and not error estimates.
Comparisons between simulations and the averaging appoximations indicate
that the error bounds hold on much longer time intervals.

We point out that this section extends certain results of [ES] in at least two
important ways: first, by using maps, we are able to incorporate delta
function ``kicks'' that could not be treated rigorously via the flow
methods of [ES]; second, the truncated Diophantine conditions used here
are more physically realistic and explicit than the small divisor conditions
used there (cf.\ \S 4.2.1).  Finally, we note that our maps need not be
polynomial here; this is particularly important for the weak-strong beam-beam
problem where
the perturbation is not polynomial.  (We also remind the reader of our
discussion of this section in the Introduction.)

We begin in Subsection 3.1 with a general ``kick-rotate'' model in one degree
of freedom. In Subsection 3.2 we apply the results of Subsection 3.1 to the
important case of the weak-strong beam-beam interaction, and in Subsection 3.3
we apply Theorem 3 and Proposition B to the H\'enon map.

\vfill\eject %%% pagebreak

\bigskip
\noi{\bf 3.1 The One Degree of Freedom Kick-Rotate Model}

\smallskip
In this subsection, for purposes of illustration we focus on a simple but
widely used class of beam dynamics models: the so-called one degree of freedom
``kick-rotate" models.  We note, however, that our methods may be
generalized to treat models with several degrees of freedom and at higher
order (this will be the subject of a future publication [DEVS]).

A circular accelerator (in storage mode) has a closed orbit, that is, there
 exists a unique solution of the equations of motion which has the periodicity
 of the (circular) acclerator.
A complete, three-degree-of-freedom description of single-particle beam
 dynamics involves three spatial coordinates in the co-moving (Frenet-Serret)
 system defined by the projection of the closed orbit on configuration space,
 and their three conjugate momenta.
It is convenient to study the dynamics in terms of a Poincar\'e map (one-turn
 map) at a fixed azimuthal location in the ring.
Here we consider one transverse degree of freedom and let $w_1$ and $w_2$
 denote the spatial coordinate and conjugate momentum in the Poincar\'e
 section.
The base-model consists of a ``rotation with unperturbed tune $\nu$''
 representing the linear ``betatron motion.''
Perturbations of this model often consist of an instantaneous change
 in momentum $w_2$ at a fixed location in the ring,
 which depends only on the spatial coordinate $w_1$ (a ``kick-map'').
If we take this fixed location to be the azimuthal position of the Poincar\'e
 section, then the perturbed dynamics is given by the so-called ``kick-rotate''
 model
$$
 w_{n+1}
 \ = \
 R \ w_n \ + \ \eps \ R \left({0\atop -H'(w_{1,n})}\right)\,,
\qquad
{\rm where}
\quad
 R \,:=\, e^{{\cal J}2\pi\nu},
 \eqno(3.1)
$$\vn
 that is, a kick followed by a rotation through the angle $2\pi\nu$ about the
 origin.
Here ${\cal J}:=\pmatrix{0 & 1 \cr -1 & 0 \cr}$ is the unit symplectic
matrix and $H'$ is the ``kick function."
Since $R$ depends only on the fractional part of $\nu$ we shall assume
 $\nu\in[0,1]$ in the following.
The map defined by Eq.\ (3.1) is symplectic since it is the composition of
 symplectic maps.
The notation $w_{1,n}$ indicates the first component of the vector
 $w_n=(w_1,w_2)_n\transp$ (we hope the reader will forgive us the ambiguity
 of using $w_n$ to denote a vector and $w_1$ or $w_{1,n}$ its first component,
 and $w_2$ or $w_{2,n}$ its second component;
 the meaning should be clear from context, since we rarely explicitly
 set $n=1$ or $n=2$).

For $R=1$, i.e.\ $\nu\in\{0,1\}$, Eq.\ (3.1) is easily solved and gives
 $w_n=(w_{1,0},-nH'(w_{1,0}))\transp$ and thus $|w_{2,n}|$ is monotonically
 increasing to infinity.
For $R=-1$ (i.e., $\nu=1/2$),
 $w_{2n}=(w_{1,0},-2nH'(w_{1,0}))\transp$ and the motion is again unbounded.
Thus for $\nu\in\{0,1/2,1\}$ and for all initial conditions where
 $H'(w_{1,0})\ne0$, the distance from the origin is monotonically increasing.
The basic question is, What happens for general $\nu$?
We shall apply the results of Section 2 to answer this question for most $\nu$
 in $[0,1]$.

Eq.\ (3.1) may be written as
$$
 w_{n+1} = R\,w_n + \eps\,R\,F(w_n) \eqno(3.2)
$$\vn
 and the transformation $ w_n = R^n\,x_n$
recasts Eq.\ (3.2) as:
$$
 x_{n+1} = x_n + \eps\,R^{-n}\,F(R^n\,x_n)
         =: x_n+\eps\,f(x_n,n\nu)\;,  \eqno(3.3)
$$\vn
which is in the standard form for averaging (cf.\ Eq.~(1.4)).

It is easy to see that
%$$
% f(x,\theta) = H'(x_1\cos 2\pi\theta + x_2\sin 2\pi\theta)
%               \pmatrix{ \sin 2\pi\theta \cr -\cos 2\pi\theta}
%             = \pmatrix{ {\partial H \over \partial x_2} \cr
%                        -{\partial H \over \partial x_1} } \;.
%$$
$ f(x,\theta) = H'(x_1\cos 2\pi\theta + x_2\sin 2\pi\theta)\;
                (\sin 2\pi\theta, -\cos 2\pi\theta)\transp
              = ( \partial H/\partial x_2\,,\,
                 -\partial H/\partial x_1     )\transp$.
Thus if we define
${\cal H}(x,\theta) := H(x_1\cos 2\pi\theta + x_2\sin 2\pi\theta)$,
 then Eq.\ (3.3) becomes
$$
 x_{n+1} = x_n + \eps \, {\cal J}\; \nabla_x {\cal H}(x_n,n\nu)\;.
 \eqno(3.4)
$$\vn
Equations (3.3) and (3.4) also define symplectic maps, since the
transformation is symplectic.

\medskip
\noi{\bf 3.1.1  The kick-rotate model in the far-from-low-order-resonance
case}

\smallskip
In this subsection, we examine the behavior of the kick-rotate
 model (3.1) in the case where the tune belongs to the
 $\eps$-dependent truncated Diophantine set $\D(\eps^\lambda\phi,R_\eps)$.
In physical terms, this means that the tune is ``far from low-order resonance."

The most useful form of ${\cal H}$ in Eq.\ (3.4) is given in terms of the
Fourier series
$
 H(\sqrt{2J}\sin 2\pi t) =\hfill\break \sum_{k\in\Z} H_k(J)\,e^{i2\pi kt},
$
from which it follows that
${\cal H}(x,n\nu) = \sum_{k\in\Z} H_k(J(x))\,e^{i2\pi k(\Phi(x)+n\nu)}$,
 where $\Phi$ and $J$ are defined by
 $x_1=\sqrt{2J}\sin(2\pi\Phi)$ and $x_2=\sqrt{2J}\cos(2\pi\Phi)$.
The averaged problem is then
$$
 y_{n+1} = y_n + \eps\,{\cal J}\,\nabla_y H_0(J(y_n))\,,
 \eqno(3.5)
$$\vn
 where
$ H_0(J) = \int_0^1 H(\sqrt{2J}\sin 2\pi t) \,dt$.
The associated (scaled) flow problem is
$$
 {d\yhat\over dt} = 2\pi\,\omega(J(\yhat))\;{\cal J} \yhat
 \;,\qquad
  \yhat(0)=x_0
 \eqno(3.6)
$$\vn
 where $2\pi\omega(J)=H_0'(J)$.
We note that the map defined in Eq.\ (3.5) is only symplectic through
$O(\eps)$;
 however, the vector field in Eq.\ (3.6) is Hamiltonian with Hamiltonian
 $H(J(\yhat))$.
It is easy to check that $J(\yhat)={1\over2}(\yhat_1^2+\yhat_2^2)$
 is constant along orbits so that $J(\yhat)=J_0=J(x_0)$ and thus
$ \yhat(t) = e^{{\cal J}2\pi\omega(J_0)t}\,x_0$.
Finally, Theorem 2 together with Propositions A and B give
$$
 w_n = e^{{\cal J}2\pi n(\nu+\eps\omega(J_0))}\,x_0 + O(\eps^{1-\lambda})
 \eqno(3.7)
$$\vn
 for $0\le n\le T/\eps$, with $\eps$ suitably restricted as in
 Proposition B for non-compactly supported perturbations,
 with $\lambda\in[0,1)$, $\nu\in\D(\eps^\lambda\phi,R_\eps)$,
 and with $R_\eps$ defined by the condition
$$
 \sum_{|k|>R_\eps}
   \| H_k'(J)\nabla_xJ \|_{D(\delta)}
 + \| H_k(J)2\pi \nabla_x\Phi \|_{D(\delta)}  \;<\; \CT\eps\,,
 \eqno(3.8) %%% (3.8) not referenced
$$\vn
 where $D(\delta)$ is the $\delta$-tube around the solution of Eq.\ (3.6)
 (see the definition of the $\delta$-tube in \S4.3.2).

%%%\vfill\eject %%% pagebreak
\medskip
\noi{\bf 3.1.2  The kick-rotate model in the near-to-low-order-resonance case}

\smallskip
For $\nu$ near low-order resonance, we write $\nu= {q\over p} + \eps a$ when
$p$ is not too large (more precisely, when $0<p\le R_\eps$ for suitable
 $\CT,\eps>0$ in (3.8)).  Thus using Eq.\ (1.7), our problem becomes
$$
 \pmatrix{ x_{n+1} \cr \tau_{n+1} } =
 \pmatrix{ x_n + \eps \,
           {\cal J}\,\nabla_x{\cal H}(x_n,n{q\over p}+\tau_n) \cr
          \tau_n  + \eps a}\;.
 \eqno(3.9) %%% (3.9) not referenced
$$\vn
We are now in the periodic case, with averaged Hamiltonian
$
 \Hchat(x,\tau) = (1/p)\sum_{n=0}^{p-1}
 H(x_1\cos(2\pi[n{q\over p}+\tau]) + x_2\sin(2\pi[n{q\over p}+\tau]))
$.
The averaged problem is
$ (y_{n+1},\tau_{n+1}) =
  (y_n + \eps\,{\cal J}\,\nabla_y\Hchat(y_n,\tau_n),\,
\tau_n +\eps a) $,
 with its associated scaled flow
$ (d\yhat/dt,d\tau/dt) =
  ({\cal J} \nabla_{\yhat}\Hchat(\yhat,\tau),\,a)$.
Solving for $\tau$ gives
$
 {d\yhat\over dt} = {\cal J}\, \nabla_{\yhat}\Hchat(\yhat,at)
$.
Theorem 1 with Propositions B and C then give
$$
 w_n = e^{{\cal J}2\pi n\nu}\,x_n
     = e^{{\cal J}2\pi n({q\over p}+\eps a)}\,\yhat(\eps n) + O(\eps)
   \eqno(3.10)
$$\vn
 for $0\le n \le T/\eps$ and for $\nu={q\over p} +\eps a$.
 However, it is not clear we have achieved a great simplification and so
 we look more closely.  It turns out that
$\Hchat\left(\exp(-{\cal J}2\pi\theta')\yhat,\theta\right) =
 \Hchat(\yhat,\theta-\theta')$,
 which suggests that an autonomous Hamiltonian system might be found with the
 symplectic transformation $\yhat\mapsto\zhat$ defined by
$\yhat=e^{-{\cal J}2\pi at}\,\zhat$.
This is indeed true and gives the autonomous system
$$
 {d\zhat\over dt} =  2\pi a\, {\cal J}\, \zhat
                   + {\cal J}\,\nabla_{\zhat}\Hchat(\zhat,0)
 \eqno(3.11)
$$\vn
 with Hamiltonian $\K(\zhat) = 2\pi a J(\zhat) + \Hchat(\zhat,0)$.
Equation (3.10) thus becomes
$$
 w_n = e^{{\cal J}2\pi n{q\over p}}\,\zhat(\eps n) + O(\eps)
 \eqno(3.12)
$$\vn
from which the behavior of the approximation is now quite transparent.

\medskip
\noi{\bf 3.1.3  Summary of the kick-rotate model}

\smallskip
We now have the following picture of the solutions of Eq.\ (3.1) on
$O(1/\eps)$ time intervals.  For $\nu\in\D(\eps^\lambda\phi,R_\eps)$ the
motion is given by Eq.\ (3.7) and thus our kick-rotate map behaves like a
twist map with tune $\nu+\eps\omega(J_0)$.  For these $\nu$ the effect of
the perturbation is slight; the up and down kicks on the integral curves
essentially cancel and the main effect of the perturbation is to create an
amplitude-dependent tune. For $\nu={q\over p}+\eps a$, we see that in the
$p$-periodic Poincar\'e map, the approximate motion moves slowly along the
phase curves given by the level curves of $\K(\zhat)$.  We thus have an
essentially complete picture of the motion (except for small gaps in $\nu$ as
discussed in Remark 2.5).

\vfill\eject %%% pagebreak

\bigskip
\noi{\bf 3.2 \ The Weak-Strong Beam-Beam Effect}

\smallskip
As a concrete example, we study the weak-strong beam-beam effect for round
 Gaussian beams in collider rings.
We treat the lattice (the sequence of transport maps through
 the various components of the accelerator)
 as a stable, linear symplectic map, and the beam-beam interaction as
 localized at the point of the ring where the bunches collide (the
 ``interaction point").
The phase space distribution of the strong beam at the interaction point
 is assumed to be stationary; in particular the beam-beam effect of the weak
 beam on the strong beam is ignored.
Therefore the beam-beam effect on the particle trajectories of the weak beam
 may be treated in the single particle picture, i.e., as a nonlinear kick
 due to the electromagnetic forces experienced while passing through a
 (longitudinally) short and time-independent external charge distribution.
We ignore coupling to the longitudinal motion, and we assume that the
 strong beam is represented by an axially symmetric charge distribution
 around the common closed orbit of the two beams in the transverse
 coordinate plane, so that it suffices to study a single phase plane.
We start by stating the model in the so-called canonical accelerator
 coordinates $v\equiv(v_1,v_2)\transp$, where $v_1$ has the dimension of a
 length and $v_2:=p_{v_1}/p_0$ is dimensionless ($p_0$ is the
 constant longitudinal momentum of the particle on the closed orbit,
 usually much larger than $p_{v_1}$, the canonical conjugate of $v_1$).
Normally the lattice is chosen so that the unperturbed beam envelope
 at the interaction point has a local minimum, and thus the linear
 lattice is represented by
$
 M:=\pmatrix{
    \cos(2\pi Q_0)       & \beta\sin(2\pi Q_0) \cr
   -\sin(2\pi Q_0)/\beta &      \cos(2\pi Q_0) }
$,
 where $Q_0\in\R$ and $\beta>0$ are the unperturbed tune and the
 unperturbed beta-function of the weak beam at the interaction point,
 respectively (the beam envelope has width of order $\sqrt{\beta}$).
The beam-beam kick is given by $v_2 \mapsto v_2  -\eta K(v_1)$,
 where $\eta:=8\pi{\sigma_1^{\ast2}\over\beta}\xi$, and where
$
 K(v_1):=
  {1 \over v_1}
  \left(1-\exp\left(-{v_1^2\over2\sigma_1^{\ast2}}\right)\right)
$.
Here  $\sigma_1^\ast$ is the spatial standard deviation of the Gaussian
representing the strong beam, and $\xi$ is the (typically small) linear
beam-beam tune shift parameter.
Our difference equation in the accelerator coordinates now
reads $ v_{n+1} = M\, v_n + \eta \,M\, ( 0, -K(v_{1,n}))\transp$.

\smallskip
\noi{\bf Remark 3.1} \  In the special case of two matched,
 axially symmetric Gaussian beams, $\xi$ is given by
 $\xi=\pm N^\ast r_p \beta/(4\pi\gamma\sigma_1^2)$,
 where $N^\ast$ is the number of particles in the strong beam,
 $r_p$ is the so-called classical particle radius of the species,
 $\sigma_1$ is the spatial RMS beam width of the two beams, and
 $\gamma>1$ is the Lorentz factor of the weak beam.

\medskip
We now rescale the variables according to
$ w \equiv (w_1, w_2)\transp
 := ( v_1/\sigma_1,\, \beta v_2/\sigma_1 )\transp$,
 where $\sigma_1$ is the standard deviation of $v_1$ for the weak beam
 when matched to its unperturbed lattice (i.e., when
 the phase space density depends only on $v\transp\,B^{-1}\,v$,
 where $B:={\rm diag}(\beta,1/\beta)$ is the beam matrix at the interaction
 point; note that $\sigma_2:=\sigma_1/\beta$ is then the standard deviation of
 $v_2$ for the weak beam).
In the rescaled variables the difference equation becomes
$$
 w_{n+1}  = R\; w_n +\eps \; R\, { 0 \choose -H'(w_{1,n})}\,,
\eqno(3.13)
$$\vn
 where $R:=e^{{\cal J}2\pi Q_0}$, $\eps:=8\pi r^2 \xi$,
 $r:={\sigma_1^\ast/\sigma_1}$, and
$
\disp
 H'(w_1):=
{1\over w_1}\left(1-\exp\left(-{w_1^2 \over
2r^2}\right)\right)$.
Thus Eq.\ (3.13) has the form of Eq.\ (3.1).
We note that $\eps$ is dimensionless and small whenever $\xi$ is
small,
 that $w_1$ and $w_2$ are dimensionless and $O(1)$ for a typical
particle trajectory of the weak beam,
 and that in a collider the two beams are typically matched to
each other so that $r \approx 1$.
By using the substitution $s^2/(2r^2)=w_1^2/(2r^2+s')$ one can show that
$$
 H(w_1)\;:=\;\int_0^{w_1}\left(1-\exp\left(-{s^2\over 2r^2}\right)\right)
          {ds\over s}
        \;=\; {1\over 2}\int_0^{\infty}
          \left(1-\exp\left( -{w_1^2 \over 2r^2+s'} \right)\right)
          {ds'\over 2r^2+s' }\;,
 \eqno(3.14)
$$\vn
 where we have taken $H(0)=0$.

Before proceeding we check the linearized behavior about the
equilibrium $w=0$.
The linearization of Eq.\ (3.13) is \
$
 w_{n+1} = G\; w_n\;,\
 G:=R\;\pmatrix{ 1 & 0 \cr - 4\pi\xi & 1 }
$,
 where we have used the fact that $H''(0)=(2r^2)^{-1}$.
The system is linearly stable if and only if $|{\rm tr}\,G|<2$, i.e.,
provided the linearly perturbed tune $Q$, defined by
 $\cos(2\pi Q) := {1 \over 2}{\rm tr}\,G =
  \cos(2\pi Q_0)-2\pi\xi\,\sin(2\pi Q_0)$, is real and satisfies
 $|\cos(2\pi Q)|<1$.
It follows that $ Q = Q_0 + \xi + O(\xi^2)$,
 thus justifying the name ``linear beam-beam tune shift parameter'' for
 $\xi$.
For $Q_0\in\{0,1/2,1\}$ we see that $|{\rm tr}\,G|=2$, consistent with the
 discussion in the paragraph immediately following Eq.~(3.1).
For $Q_0\in\{1/4,3/4\}$, $|{\rm tr}\,G|=2\pi|\xi|$ and thus we have linear
 stability, which is consistent with the results of Subsection 3.2.2.

\vfill\eject %%% pagebreak

\medskip
\noi{\bf 3.2.1 \ The weak-strong beam-beam effect in the
far-from-low-order-resonance case}

\smallskip
For $Q_0\in\D(\eps^\lambda\phi,R_\eps)$ the motion is given by Eq.\ (3.7),
where $\nu\equiv Q_0$, and where $\omega$ is determined as follows.
We use Eq.\ (3.14) to obtain
$
 H_0(J)\,:=\,\int_0^1 H(\sqrt{2J}\sin(2\pi t))\,dt \,=\, {1\over2}
 \int_0^{J/(2r^2)} \left(1-e^{-w} I_0(w)\right) {dw\over w}
$,
  where $I_0$ is the zero-th order modified Bessel function and where we
have used the expansion
 $\exp(x\cos(y))=I_0(x)+2\sum_{k=1}^{\infty} I_k(x)\cos(ky)$.
Omega is given by
$$
 2\pi\omega(J) \,:=\, H_0'(J) \,=\,
 {1 \over 2J} \left(1-\exp\left(-{J\over 2r^2}\right)
        I_0\left({J \over 2r^2}\right)\right)
 \,=\, {1\over 4\pi J} \int_0^{2\pi}
  \left(1-\exp\left(-{J\sin^2\vartheta\over r^2}\right)\right)\,d\vartheta\,.
 \eqno(3.15)
$$\vn
The amplitude-dependent tune shift $\eps\omega(J_0)$ is identical to that
derived in [ES] and justifies the use of the delta function there.
Notice also that $\eps\omega(0)=\xi$, in agreement with the linearization
above.

%%%\vfill\eject %%% pagebreak
\medskip
\noi{\bf 3.2.2 \ The weak-strong beam-beam effect in the
near-to-low-order-resonance case}

\smallskip
In Subsection 3.1.2 we found the Hamiltonian for the autonomous system (3.11)
 to be $\K(\zhat) = 2\pi aJ(\zhat)+ \Hchat(\zhat,0)$, where
 $\Hchat(\zhat,0)=
  (1/p)\sum_{n=0}^{p-1}H(\zhat_1\cos[2\pi nq/p]+\zhat_2\sin[2\pi nq/p])$
 and $Q_0=q/p+a\eps$.
Since $H(x)$ approaches zero for large $x$, $\K(\zhat)$ approaches
 $2\pi aJ(\zhat)$, and for $a\not=0$ the integral curves become circles
at large distances from the origin.
The motion on these circles is clockwise for positive $a$ and counterclockwise
 for negative $a$, thus a bifurcation in the phase plane portrait
 occurs at $a=0$.
In the case where $q/p \in \{0,1/2,1\}$ it is easy to see that
 $\Hchat(\zhat,0)=H(\zhat_1)$, and for $q/p \in \{1/4,3/4\}$ one also easily
 finds
 $\Hchat(\zhat,0)=1/2\,[H(\zhat_1)+H(\zhat_2)]$ since $H$ is an even function.
For $q/p \in \{1/3,2/3\}$ we find $\Hchat(\zhat,0)=1/3\,
 [H(\zhat_1)+H(-\zhat_1/2+\sqrt{3}\zhat_2/2)+H(-\zhat_1/2-\sqrt{3}\zhat_2/2)]$.
We briefly discuss the phase plane portraits for $\K$ in these cases
(see [DEV] for more figures).

In the first case ($q/p\in\{0,1\}$) and for $a=0$ we have $d\zhat_1/dt=0$ and
 $d\zhat_2/dt=H'(\zhat_{1,0})$.
Thus the motion is identical to the exact case, as discussed just before
Eq.\ (3.2), since Eqs.\ (3.1) and (3.3) and the associated averaged problem are
identical. For $a$ small but positive, the origin is a
 (nonlinearly) stable center and the phase portrait is a one-parameter
 family of ovals which are long and thin in the $\zhat_2$ and $\zhat_1$
 directions respectively.
As $a$ increases to modest values the ovals become circular, consistent with
 the expectation of ``stability far from low-order resonance.''
As $a$ decreases from zero, the origin becomes a saddle, and two centers
emerge from infinity at $(\pm c,0)$, where $c\sim 1/\sqrt{2\pi|a|}$
for $|a|$ small.
As $a$ decreases further, the centers coalesce with the saddle at
$4\pi a r^2=-1$, and for $4\pi a r^2<-1$ the only critical point is
a center at the origin, again consistent with our expectation of stability
 (see Figure 1).

The motion for $q/p=1/2$ in the period two Poincar\'e map is identical with
 the motion for $q/p=1$; the intermediate values may be obtained by
 rotating the phase plane portrait by a half turn (also see Figure 1).

\medskip
\centerline{\epsfxsize=14cm\epsfbox{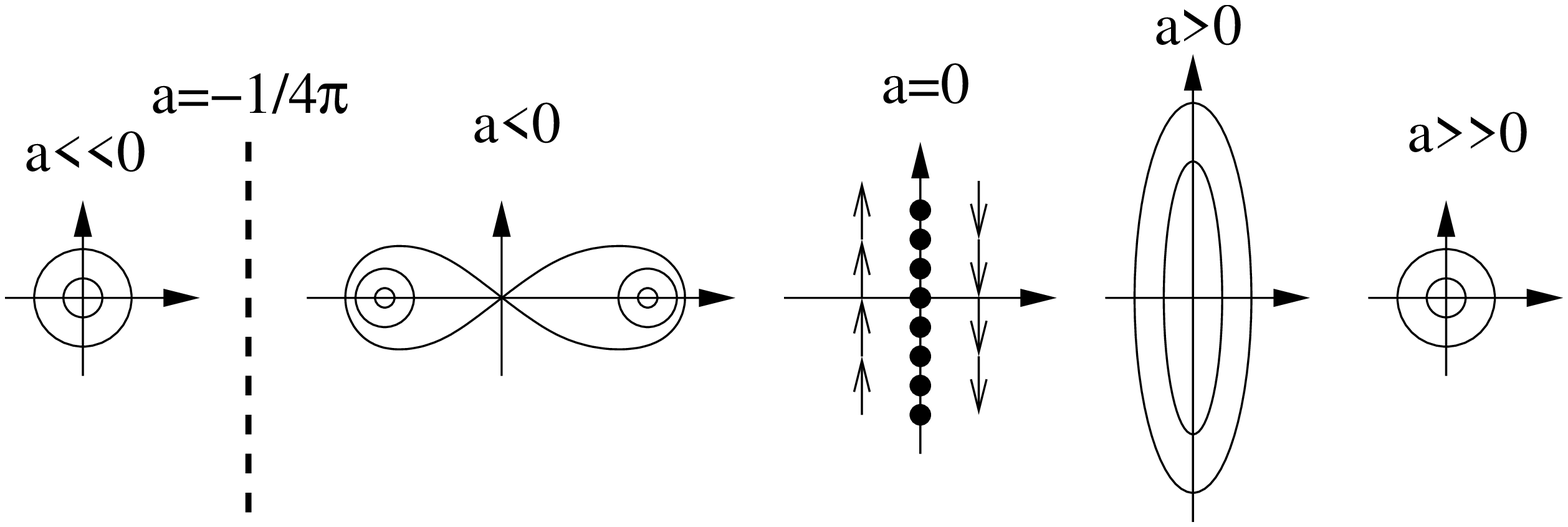}}
\centerline{{\bf Figure 1:} The qualitative phase plane portraits for
 $r=1$ in the case $q/p\in\{0,1/2,1\}$}
\medskip

For $q/p \in \{1/4,3/4\}$ the phase plane portrait (see Figure 2)
 has a four-fold symmetry, being invariant under reflections about
 the two axes and about the lines $\zhat_2=\pm \zhat_1$.
The origin is a critical point and its linearized vector field has eigenvalues
 $\pm 2\pi i (a-a_c)$, where $a_c=-1/(8\pi r^2)$.
Thus the origin is a (nonlinearly) stable center for $a \ne a_c$, and it is
 easily checked that the origin is also a stable center for $a=a_c$ and that
 the rotation is clockwise for $a > a_c$ and counterclockwise for $a \le a_c$.
For $a\ge0$ there are no other equilibria and the phase plane portrait is
 a one-parameter family of concentric ovals.
For $a$ small the (closed) integral curves look like four-pointed stars,
 with smoothed points on the
 axes, and as $a$ increases the curves become circles.
For $ a_c < a < 0$ there are eight nonzero critical points.
The four critical points $(\pm c,\pm c)$ are centers and the four at
 $(0,\pm c)$ and $(\pm c, 0)$ are saddle points, where $c$ is the unique
 positive root of $4\pi a c + H'(c) = 0$.
The critical points form an island structure in a neighborhood of radius $c$
of the origin in the phase plane.
This island structure emerges from infinity as $a$ decreases through zero and
 coalesces into the origin as $a$ decreases to $a_c$.
For $a \le a_c$, the origin is again the only equilibrium, and it is a stable
 center with counterclockwise rotation.
The portrait is again a one-parameter family of ovals approaching circles as
 $a$ decreases from $a_c$.

\medskip
\centerline{
 \vbox{\hbox{\hskip2mm $a=-0.050$}\hbox{\epsfxsize=2.25cm\epsfbox{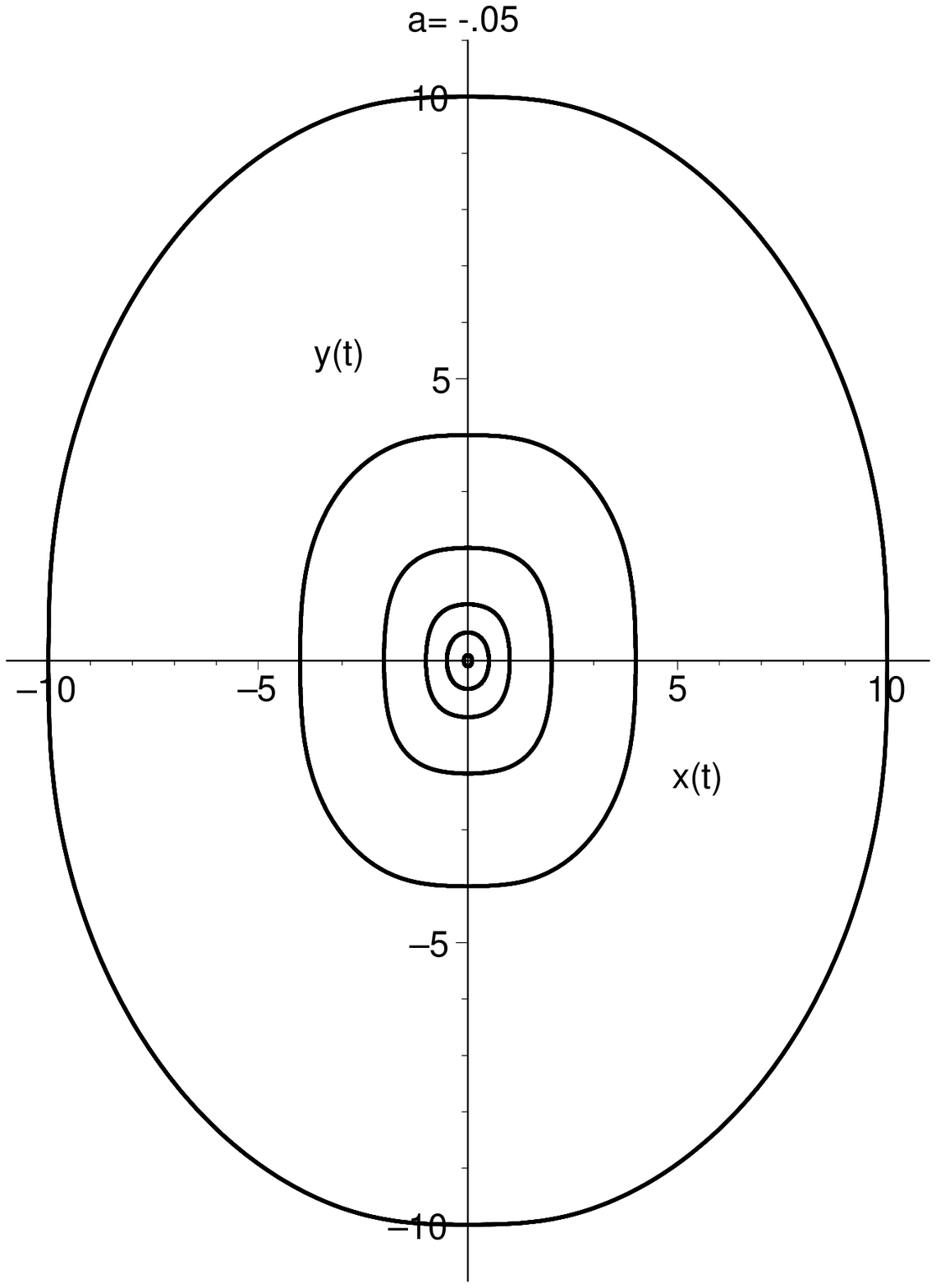}}}
 \vbox{\hbox{\hskip2mm $a=-\pi/8$}\hbox{\hskip8mm\epsfxsize=0.065cm\epsfbox{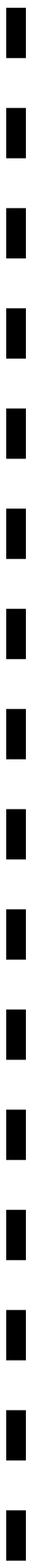}}}
 \vbox{\hbox{\hskip2mm $a=-0.005$}\hbox{\epsfxsize=2.25cm\epsfbox{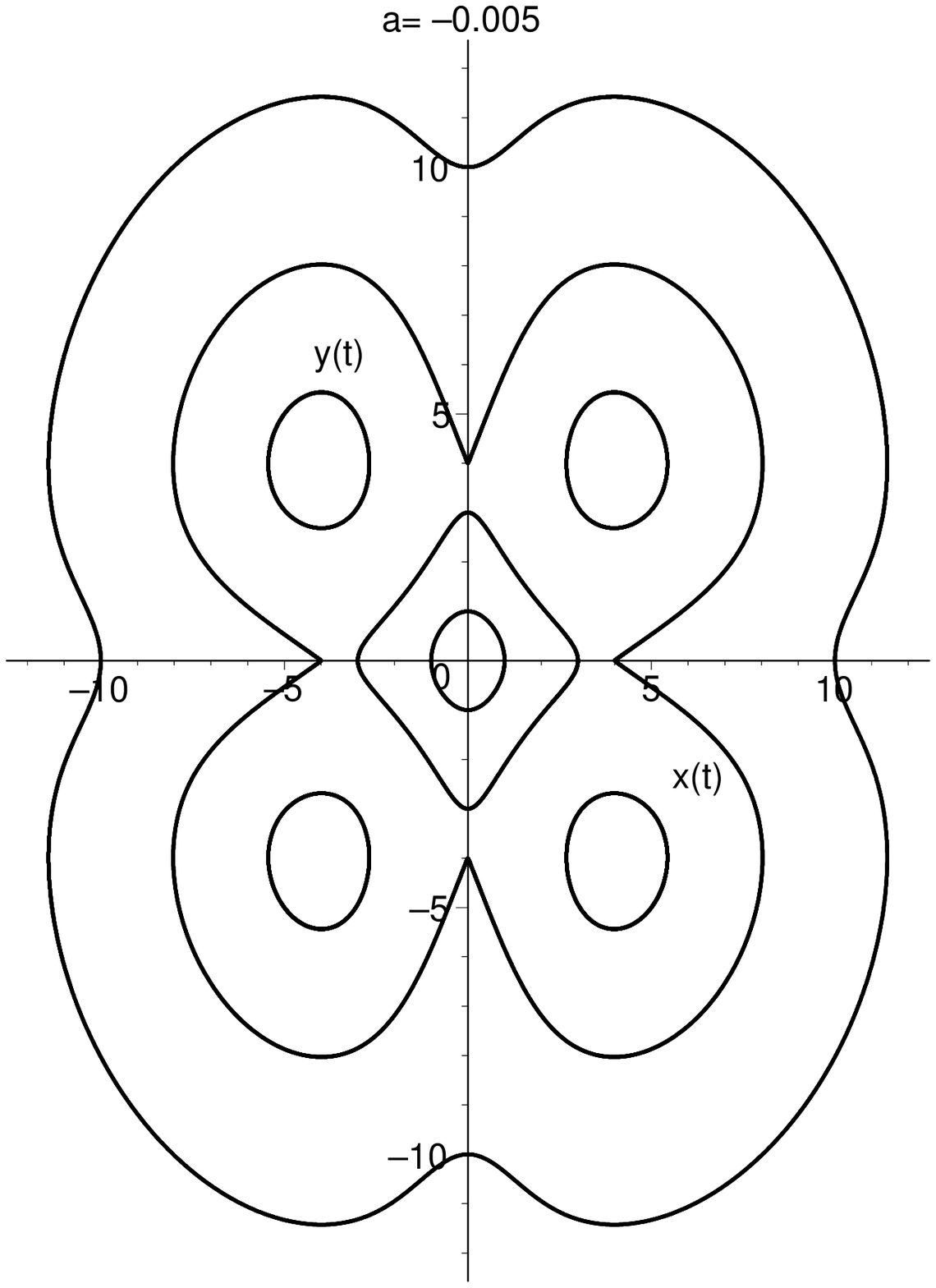}}}
 \vbox{\hbox{\hskip2mm $a=-0.000$}\hbox{\epsfxsize=2.25cm\epsfbox{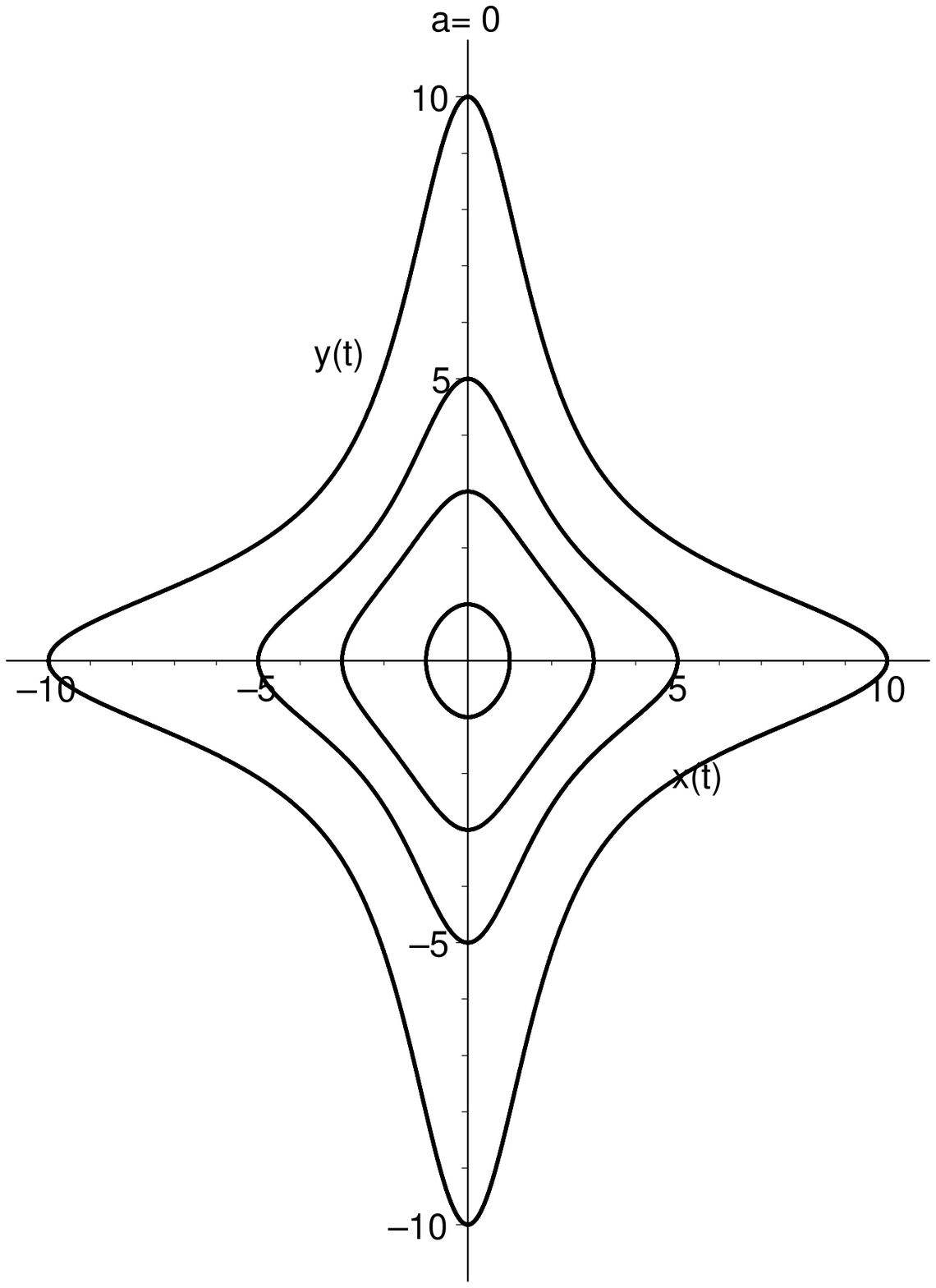}}}
 \vbox{\hbox{\hskip2mm $a=+0.005$}\hbox{\epsfxsize=2.25cm\epsfbox{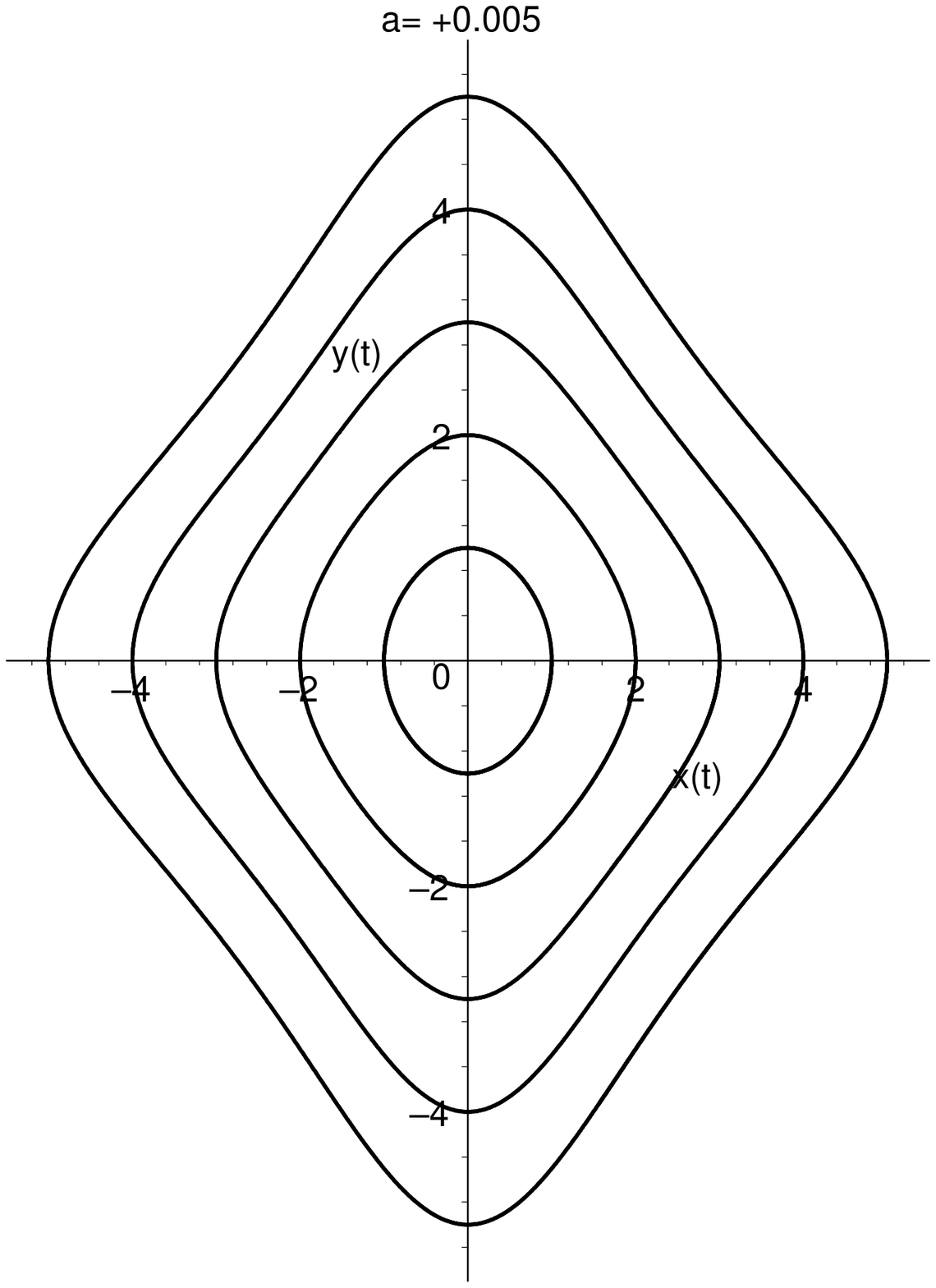}}}
 \vbox{\hbox{\hskip2mm $a=+0.050$}\hbox{\epsfxsize=2.25cm\epsfbox{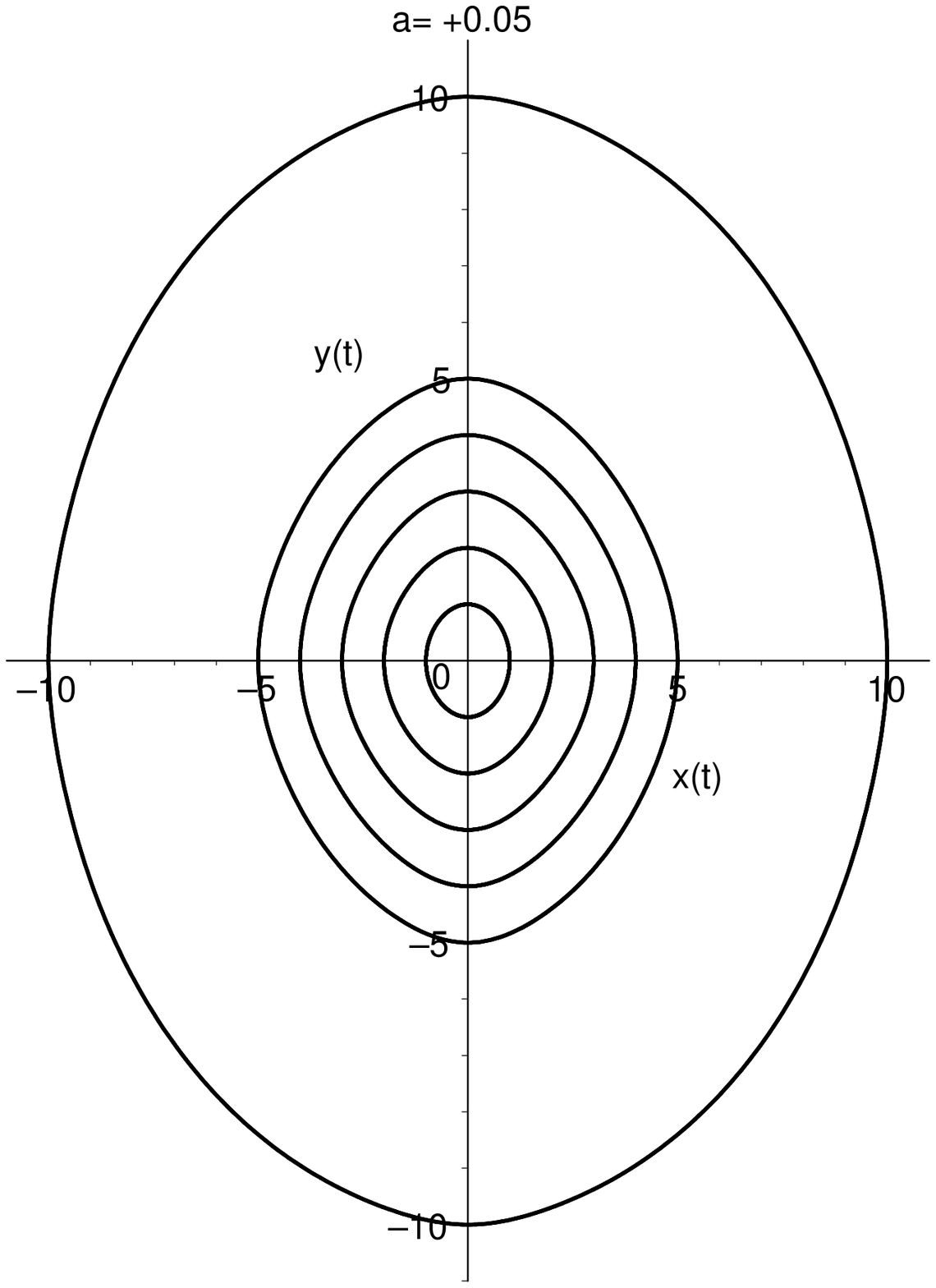}}}
}
\centerline{{\bf Figure 2:} The phase plane portraits for $r=1$ in the case
  $q/p\in\{1/4,3/4\}$.}
%\centerline{Left to right:
% $a=-0.05$,
%% $a=-0.035$,
%% $a=-0.025$,
% $a=-0.005$, $a=0$, $a=+0.005$, $a=+0.05$.}
\medskip

Because $H$ is an even function, $\Hchat$ is the same for all
 $q/p\in\{1/6,1/3,2/3,5/6\}$.
Thus the phase plane portraits are the same for resonances of order three and
six, and these portraits have a six-fold symmetry, being invariant under
reflections about the axes $\zhat_1=0$, $\zhat_2=0$ and the lines
$\zhat_2=\pm\zhat_1/\sqrt{3}$ and $\zhat_2=\pm\sqrt{3}\,\zhat_1$.
Qualitatively, the behavior as a function of $a$ is similar to that in the case
of resonance of order four (e.g., the island structure is similar, but
there are now six rather than four islands).
The critical value $a_c$ at which the islands coalesce in the origin
 turns out to be the same as in the case $p=4$.

\medskip
\noi{\bf 3.2.3 \ Summary of the weak-strong beam-beam effect}

\smallskip
Our basic equation is Eq.\ (3.13) with $R$ and $H'$ defined there.
Remark 2.5 and the summary in Subsection 3.1.3 apply.
Here we emphasize that the motion depends only on $\xi$
(or equivalently $\eps$), and on the fractional part of $Q_0$, and
that we have a fairly complete description for $Q_0\in[0,1]$ over
$O(1/\xi)$ time intervals.
Away from low-order resonances, the motion is given by Eq.\ (3.7), with
$\nu=Q_0$ and with $\omega$ defined in Eq.\ (3.15).
Thus the motion takes place approximately on circles with an
amplitude-dependent tune.  Near low-order resonances the behavior is given by
Eq.\ (3.12), and $\zhat(t)$ evolves according to the time-independent
Hamiltonian $\K(\zhat):=2\pi aJ(\zhat)+(1/p)\,\sum_{n=0}^{p-1}
  H(\zhat_1\cos(2\pi nq/p)+\zhat_2\sin(2\pi nq/p))$.
As described above in Subsection 3.2.2, this Hamiltonian has a rich variety
of behaviors depending on the order $p$ of the resonance, and on the
displacement
$a\eps$ from the resonance.  In particular the behavior varies considerably
for $a>0$, $a=0$ and $a<0$. Finally, we again emphasize that while our
description is fairly complete, there are gaps between the regions of
validity of the nonresonant normal form which does not depend on $Q_0$, and
the resonant normal form which does depend on $Q_0$ (cf.\ Remark 2.5).

\bigskip
\noi{\bf 3.3 \ The H\'enon Map}

\smallskip
We now apply Theorem 3 to the H\'enon map
(in beam dynamics this map is a standard model for the
effect of a localized sextupole magnet in an otherwise linear lattice).
The standard form of the H\'enon map is Eq.\ (3.1) with
$H(w_1) = w_1^3/3$.  This gives Eq.\ (3.4) with
${\cal H}(x,\theta) = (x_1\cos2\pi\theta + x_2\sin2\pi\theta)^3/3$,
which clearly has zero average.
It follows that $f(x,\theta)={\cal J}\,\nabla_x{\cal H}(x,\theta)$
in Eq.\ (1.4) has zero average, so that hypothesis (jw) of Theorem 3
is satisfied.
Thus, by Theorem 3 and Proposition B, for appropriate $\eps,T>0$,
$\nu\in D(\phi,R_\eps)$, and
for any $0\le\alpha\le1$, we have $|x_n-x_0|=O(\eps^\alpha)$
on the discrete time interval $0\le n\le T/\eps^{2-\alpha}$.

\smallskip
\noi{\bf Remark 3.2} \ The above discussion simply applies Theorem 3 as is
(and thus also covers the case of more general $H$), but when $\cal H$ has
a finite Fourier series (e.g.\ when $H$ is a polynomial, as above) the proof
of Theorem 3 may be simplified, both in terms of the smoothness requirement
(see Remark 4.4) and in terms of the estimates in Lemma 2.  In particular,
for the H\'enon map above, $g_k=0$ except for $|k|\in\{1,3\}$, so taking
$R_\eps=3$, we see that the series defining $C_1$ and $C_2$ in Lemma 2 have
only four terms each, while the tail-series of Lemma 2 vanishes.

\bigskip\medskip
\noi{\bf 4. \ Proofs and Additional Mathematical Results}

\medskip\noi
As the title indicates, this is the most mathematical section of the
paper.  Subsection 4.1 treats periodic maps; this is quite straightforward,
and may be read as a kind of introduction to the deeper results of the next
subsection. Subsection 4.2 concerns the considerably more complex case of
maps far from low-order resonance, and requires a (short) discussion of
small divisors and truncated Diophantine conditions.  The use of such
conditions is not new (for example, related conditions are used to obtain
general multiphase averaging results in [ABG]), but as explained in the
introduction, we believe our use of them in the present context is the most
innovative aspect of this paper from the viewpoint of applied mathematics.

\vfill\eject %%% pagebreak

\bigskip
\noi{\bf 4.1 Periodic Systems}

\smallskip
In this subsection we give a self-contained presentation of the
remarkably simple technology required to prove the averaging
principle for maps with periodic perturbations.  This consists of the
Besjes inequality for periodic functions (below), followed by its
application to the proof of Theorem 1.

\medskip
\noi{\bf 4.1.1 The Besjes inequality for periodic functions}

\smallskip
Let $U\subset\Rn$ be open, and $S=U\times\N$.  The Besjes inequality
relies in an
essential way upon the following assumption concerning the function
$g:S\to\R$, periodic with period $p$ in its second argument:

\smallskip
(iv) For each $x\in U$, \ $\;\disp\sum_{n=0}^{p-1} g(x,n) = 0$

\medskip
When $g$ has period $p$ in $n$ and satisfies (iv), we say it has {\sl
zero mean in} $n$.  We now state the Besjes inequality for periodic
maps as

\proclaim Lemma 1.
Let $U\subset\Rn$ be open, $S=U\times\N$, and suppose $g:S\to\R$
satisfies assumptions (i), (ii) (from \S 2.1)
and (iv) above and is globally $x$-Lipschitz with Lipschitz constant
$L\ge0$.  If $\{x_n\}_{n=0}^\infty\subset U$ is a sequence for
which the successive differences $x_{n+1}-x_n$ are bounded by $M$
(i.e., $\sup_n|x_{n+1}-x_n|\le M$), then for all $N\in\N$,
$$
 \Bigl|\sum_{n=0}^{N-1} g(x_n,n)\Bigr|
 \ \le \
 {1\over2}\,NpLM
 \,+\,
  p\,\|g\|_S.
$$

\noi{\it Proof.} Using the notation $[a]$ to designate the greatest
integer in $a$, we first set $l = [(N-1)/p]$ (so that $l$ is the
number of periods of $g$ contained in the segment
$\{0,1,2,\ldots,N-1\}$).  Then using the
fact that $g$ is periodic and of zero mean, we write
$$
 \sum_{n=0}^{N-1} g(x_n,n)
 \ =
 \sum_{k=0}^{l-1}
 \,\sum_{n=0}^{p-1}
 \Bigl(g(x_{n+kp},n) - g(x_{kp},n)\Bigr)
 \ + \
 \sum_{n=lp}^{N-1} g(x_n,n)\,.
$$\vn
Now since $g$ is Lipschitz in its first argument,
and since $|x_{n+kp} - x_{kp}|\le Mn$, we have
$$
 \Bigl|\sum_{n=0}^{N-1} g(x_n,n)\Bigr|
 \ \le \
 \sum_{k=0}^{l-1}
 \,\sum_{n=0}^{p-1}LMn
 \ \ + \
 \sum_{n=lp}^{N-1}|g(x_n,n)|
$$\vn
$$
 \le \
 lLM\,{p(p-1)\over2}
 \ \,+ \,\
 p\,\|g\|_S
 \ \le \ \
 {1\over2}\,NpLM
 \ + \
  p\,\|g\|_S\,.\quad//
$$

\smallskip
\noi{\bf Remark 4.1} \ The original version of this lemma (Lemma 1
of [Bes]) was formulated for use in the proof of averaging
principles for ODEs on $O(1/\eps)$ timescales, and we use its
analog in a similar way below for maps.  The original lemma bounds the time by
a constant that is $O(1/\eps)$ and gives a final bound that is $O(\eps)$,
independent of time.  We have found, however, that retaining the (here
discrete) time-dependence makes the result more versatile (cf.\ the proof of
Theorem 3 below).

\medskip
\noi{\bf Remark 4.2} \ Lemma 1 (and many of its generalizations) may
also be proved using ``summation by parts,"  as in the proof of Lemma 2
below.

\medskip
We now illustrate the use of Lemma 1 by using it to prove Theorem 1.

\medskip
\noi{\bf 4.1.2 Proof of Theorem 1}

\smallskip
Assume the hypotheses of Theorem 1 (cf.\ \S 2.1).  It is clear from
assumption (iii) that the solutions $x_N$ and $y_N$ exist uniquely for all
$N\in\N$.  To see that the approximation relation holds, we write
$$
 |x_N-y_N|
 \ = \
 \eps \,
 \biggl|
 \sum_{n=0}^{N-1}
 \Bigl(f(x_n,n) - \fhat(y_n)\Bigr)
 \biggr|
 \ = \
 \eps \,
 \biggl|
 \sum_{n=0}^{N-1}
 \Bigl(f(x_n,n) - f(y_n,n) + f(y_n,n) - \fhat(y_n)\Bigr)
 \biggr|
$$\vn
$$
  \le \
 \eps L \sum_{n=0}^{N-1}|x_n-y_n|
 \ + \
 \eps \,
  \biggl|
 \sum_{n=0}^{N-1}
 \fwig(y_n,n)
 \biggr|
$$\vn
where $L$ is the $x$-Lipschitz constant of $f$, and where
$\fwig(y,n):= f(y,n) - \fhat(y)$ (the ``oscillating part of
$f$") satisfies the hypotheses of Lemma 1 with $U=\Rn$ (in particular,
$\fwig$ has
zero mean and $y$-Lipschitz constant $2L$).  Using the fact (from
Eq.\ (1.2) and assumption (i)) that
$|y_{n+1}-y_n|\le M:=\eps\|\fhat\|_\Rn$, we have \
$\disp
 |x_N-y_N|
 \ \le \
 \eps\, L \sum_{n=0}^{N-1}|x_n-y_n|
 \ + \
 \eps\,
 {1\over2}\,N\,p\,2L\,\eps\|\fhat\|_\Rn
 \ + \
  \eps\,p\,\|\fwig\|_S$.
Thus
$\disp
|x_N-y_N|
 \ \le \
 \eps L \sum_{n=0}^{N-1}|x_n-y_n|
 \ + \
 \eps\,p\,
 \bigl(LT \|\fhat\|_\Rn + \|\fwig\|_S\bigr)
$
for $0<N\le T/\eps$.
Applying Gronwall's inequality for sequences (Lemma 3 in the
Appendix) and setting
$C = \bigl(LT \|\fhat\|_\Rn + \|\fwig\|_S\bigr) e^{LT}$ gives
$|x_N-y_N| \ \le \ Cp\,\eps$ \ for \ $0<N\le T/\eps$, as claimed.
The second part of Theorem 1 (namely $|x_n-y(n)|\le(Cp+C')\eps$ for
$0\le n\le T/\eps$) follows from Lemma 4 (Appendix) and the triangle
inequality.\quad//

\smallskip
\noi{\bf Remark 4.3} \ The preceding is no doubt one of the simplest
possible proofs of an averaging principle for maps.  Part of the
simplicity derives from the use of Lemma 1, and part derives from
the assumption of compact support (iii), which permits us to dispense
with questions of the existence intervals for solutions.  Thus, although
assumption (iii) is often invalid in practice, by using it we are able to
show that the basic estimates of the averaging method do not require
restrictions on the size of $\eps$; such restrictions are instead
introduced by considering solutions' existence intervals, or by
methods of proof which rely on near-identity transformations (which may in
turn require restrictions on $\eps$ for their inversion).  Of course our
results may be extended to cases with finite existence intervals (see
Proposition B, \S2.4), and may also be combined with more traditional
transformation methods to obtain efficient results at higher order [DESV].

\bigskip
\noi{\bf 4.2 Systems Far From Low-Order Resonance}

\smallskip
\noi In this subsection we generalize the Besjes inequality to functions
far from low-order resonance in their second argument.  We then use this
inequality to prove Theorems 2 and 3. First, however, we present the
following brief discussion.

%%%\vfill\eject %%% pagebreak
\medskip\noi
{\bf 4.2.1 Resonant zones, Diophantine conditions, and the
ultraviolet cutoff}
\smallskip

Before stating and proving our next analog of Besjes' inequality, we
discuss aspects of resonance, small divisors and Diophantine
conditions that will be needed in the sequel.  A more comprehensive
introduction may be found in [Yo].

\medskip
\noi{\sl Zone Functions and Diophantine Conditions}

\smallskip
In dynamical systems, Diophantine conditions arise naturally as a
means of ``controlling small divisors" and ``avoiding resonances."
Typically, in one dimension, divisors of the form $e^{2\pi ik\nu}-1$
(with $0\not=k\in\Z$ and $0\not=\nu\in\R$) occur as the denominators
of terms in a series indexed over $k$, together with numerators which
decrease to zero with increasing $|k|$.
Clearly divisors cannot vanish, so rational (or ``resonant") values
of $\nu$ must be avoided.  And although irrational $\nu$ do not
cause divisors to vanish, when ``nearly resonant," they may
generate such small divisors as to cause divergence of
the series in which they occur.

By using a suitably decreasing {\sl zone function} $\phi:\R_+\to\R_+$
(the inverse of which is called an ``approximation function" in
[R\"u]), we define the ``highly nonresonant" values of $\nu$
as  those belonging to the corresponding {\sl Diophantine set}
$$
 \D(\phi) = \{\nu\in\R\, \big|\, |e^{2\pi ik\nu}-1|\ge
 \phi(|k|),\
 k\in\Z\backslash\{0\}\},
 \eqno(4.1)
$$\vn
which is a Cantor set. The Diophantine set $\D(\phi)$ may be
thought of as $\R$ with countably many {\sl zones\/} removed, where
the zone ${\cal Z}_k=\{\nu\in\R\, \big|\, |e^{2\pi ik\nu}-1|< \phi(|k|)\}$
 corresponding to a particular $k\not=0$ is the countable union of open
 intervals centered on rational numbers of the form $q/k$ ($q\in\Z$).
To better see the structure of $\D(\phi)$, consider its intersection
with the interval $[0,1]$.  For each fixed $k>0$ we remove $k$
intervals of length $2\delta$ from $[0,1]$, where
$|e^{i2\pi k(\delta+l/k)}-1| = |e^{i2\pi k\delta}-1| < \phi(k)$.
For small $\phi(k)$, this gives $\delta \approx \phi(k)/(2\pi k)$,
and thus the total length of ${\cal Z}_k \cap [0,1]$ is $2\delta k
\approx \phi(k)/\pi$.  It follows that the total length of the union
$\cup_k {\cal Z}_k\cap[0,1]$ of the overlaps of all zones ${\cal Z}_k$
with $[0,1]$ is (approximately) bounded by
$\sum_k{\rm length}({\cal Z}_k\cap[0,1]) \approx (1/\pi)\int_1^\infty
\phi(k)\,dk$. Thus a typical zone function of the form $\phi(r) = \gamma
r^{-(\tau+1)}$ with $\tau>0$ removes zones of total length no more than
$\gamma/(\pi\tau)$ from $[0,1]$. When this total length is less than
one, the Diophantine set ${\cal D}(\phi)$ has positive measure (and is
therefore nonempty).

More generally, if the zone function $\phi$ decreases too
slowly, then the union of the excluded zones may be so large that its
complement, $\D(\phi)$, is empty.  Conversely, if $\phi$ decreases
too rapidly, then $\D(\phi)$ may be too large, and may contain
values of $\nu$ so close to resonance as to cause divergence of the
series in which small divisors appear.

The following terminology is useful for describing zone functions
that permit convergence of the series arising in the proof of Lemma 2
below.  If $U\subset\Rn$ is open, and $f:U\times\R\to\Rn$ has period 1 in
its second argument and Fourier series
$f(x,\theta) \sim \sum_{k\in\Z} f_k(x)\,e^{2\pi ik\theta}$
(where the $k$th Fourier coefficient is
$f_k(x)=\int_0^1f(x,\theta)\,e^{-2\pi ik\theta}\,d\theta$, requiring
only that $f$ is integrable in $\theta$), then given a zone
function $\phi$ such that $\D(\phi)\not=\emptyset$,
we say that $\phi$ is {\sl adapted to $f$ on $U$}
provided
$$
 \sum_{0\not=k\in\Z}{\|f_k\|_U\over\phi(|k|)}\, < \infty
 \qquad{\rm and}\qquad
 \sum_{0\not=k\in\Z}{\|Df_k\|_U\over\phi(|k|)}\, < \infty\,,
 \eqno(4.2)
$$\vn
where $Df_k$ denotes the derivative of the function $f_k:U\to\Rn$.
Smoothness conditions on $f$ assuring the existence of zone functions
adapted to $f$ are not severe, as we now show.

\medskip
\noi{\sl Smoothness Conditions Ensuring the Existence of Adapted Zone
Functions}

\smallskip
Several questions naturally arise concerning the relationship
between the smoothness of $f$ and the existence of zone functions
adapted to
$f$ as in Eq.\ (4.2).  Formulating the sharpest possible conditions in
this direction is somewhat delicate, but the following brief
discussion should serve as a good starting point.

We first recall that for $\tau>0$, the zone function $\phi(r) =
\gamma r^{-(\tau+1)}$ generates a nonempty Diophantine set
$\D(\phi)$ provided $\gamma>0$ is sufficiently small  (see the
 preceding discussion, or the more extensive discussion in \S1.2 of
[BHS]).
We assume that $f:U\times\R\to\Rn$ is of
class $C^{p+1}(U\times\R)$ and of compact support in the first
argument, uniformly with respect to the second (cf.\ assumption
(jjj) in \S 2.2).  Integrating the $k$th Fourier coefficient
$f_k(x)=\int_0^1f(x,\theta)\,e^{-2\pi ik\theta}\,d\theta$ by parts
$p$ times with respect to $\theta$ gives \hfill\break
$
f_k(x)=(2\pi ik)^{-p}
\int_0^1
\bigl[\partial^p f/\partial\theta^p\bigr](x,\theta)
e^{-2\pi ik\theta}
\,d\theta$.
Then taking the supremum over $x\in U$ of both sides of this
 expression gives
 $\|f_k\|_U \le C(f,p)|k|^{-p}$, where
 $C(f,p) = {1\over (2\pi)^p} \sup_{x\in U}\int_0^1\Bigl|
 {\partial^p f\over\partial\theta^p}(x,\theta)\Bigr|\,d\theta$.
The same estimate holds for $\|Df_k\|_U$ with $C(f,p)$ replaced
by
$ C'(f,p) = {1\over (2\pi)^p} \sup_{x\in U} \int_0^1\Bigl|
  {\partial^{p+1} f\over\partial x\partial\theta^p}(x,\theta)
  \Bigr|\,d\theta$.

Using these estimates, we immediately deduce that both of the series
in Eq.\ (4.2) are convergent provided that $p>\tau+2$.  Conversely,
we see that whenever $p\ge3$, there exists a zone function
$\phi(r) = \gamma r^{-(\tau+1)}$ with $0<\tau<p-2$ which generates
nonempty Diophantine sets $\D(\phi)$ (for $\gamma$ sufficiently
small) and which is adapted to $f$ in the sense of Eq.\ (4.2).
This justifies our assumption (j) in Theorems 2 and 3.

\smallskip
\noi{\bf Remark 4.4} \
A more refined (and lengthy) argument
shows that the existence of $\phi$ adapted to $f$ does not require quite
as much smoothness as we demand above; we start our discussion under
the assumption $f\in C^{p+1}(U\times\R)$ primarily for simplicity.
Of course, when $f$ has a (sufficiently short) finite Fourier series,
the decay rate of its terms is not an issue, and the smoothness
requirement may be reduced to $C^1$.

\medskip
\noi{\bf Remark 4.5} \  Although our results for system (1.4) as presented
in this paper do not apply to the case of analytic perturbations $\eps f$
(since analytic $f$ with compact support vanishes identically), it would
not be especially difficult to extend our theory to this case. For
analytic $f:U\times\To\to\R$ with Fourier coefficients $f_k$ decreasing
exponentially as, say, $\|f_k\|_U\le \Gamma e^{-\beta|k|}$, it would be
appropriate to use exponentially decreasing zone functions, for which the
preceding discussion is easily modified. In fact, given any $\rho>0$, the
zone function $\phi(r) = \gamma e^{-\rho r}$ generates nonempty
Diophantine sets $\D(\phi)$ for small enough
$\gamma>0$.  The decay rate $\beta$ of the $f_k$ must of course exceed
$\rho$, which can be arranged provided $f$ is analytic in its second
argument with {\sl analyticity parameter\/} $\alpha>\rho$ (this is an
instance of the Paley-Wiener Lemma; cf.\ [PW] or [BHS]).  Roughly
speaking, the analyticity parameter $\alpha$ is a measure of the minimum
distance by which $f$ may be extended as an analytic function of the
complex torus (see also \S 4.3.3 of [DEG] for an elementary
discussion in the two-dimensional case).

It is interesting to note that Diophantine conditions
corresponding to exponentially decaying zone functions $\phi$ may
be strictly weaker than the weakest small-divisor conditions
ordinarily used in dynamical systems, the so-called {\sl Bruno
conditions\/} (also spelled Brjuno or Bryuno; here ``strictly weaker"
means that the set
$\D(\phi)$ properly contains the set of $\nu$ subject to Bruno
conditions).  This is however not surprising, since Bruno conditions
apply to situations (such as conjugacies of circle
diffeomorphisms, or KAM theory) in which countably many series with small
divisors must simultaneously converge.  By contrast, in Lemma 2 we require the
convergence of only two series (in the language of [BHS], ours is a
``one-bite" small-divisor problem).

\medskip
\noi{\sl The Ultraviolet Cutoff and Truncated Diophantine
Conditions}
\smallskip

Finally, we introduce the notion of ultraviolet cutoff, which is
important in physical applications of Diophantine conditions.  To
understand why, note that typically in applications, the $\nu$ that
are required to be Diophantine are physical parameters.
But checking whether a given
$\nu$ belongs to a Cantor set of the form $\D(\phi)$ is a practical
impossibility, since each point of $\D(\phi)$ has points
arbitrarily
close to it that are not in $\D(\phi)$.  In other words, deciding if
$\nu$ belongs to $\D(\phi)$ requires $\nu$ to be
specified with infinite precision.  Practically of course, it is only
possible to specify physical parameters with finite precision.  We
surmount this difficulty by introducing {\sl truncated Diophantine
conditions\/} of the form
$$
\D(\phi,R) = \{\nu\in\R\, \big|\, |e^{2\pi ik\nu}-1|\ge
 \phi(|k|),\
 k\in\Z \ {\rm with} \ 0<|k|\le R \}.
 \eqno(4.3)
$$\vn
When $\nu\in\D(\phi,R)$, we say $\nu$ is {\sl Diophantine to order
$R$ with respect to $\phi$}, and we call $R$ the {\sl truncation
order\/} or ({\sl ultraviolet}) {\sl cutoff}.  Note that $\D(\phi,R)$ is
an approximating superset of $\D(\phi)$ with nonempty interior which
converges to $\D(\phi)$ as $R\to\infty$. To decide whether $\nu$
belongs to $\D(\phi,R)$, one checks only finitely many inequalities.

As a rough general rule, results in dynamical systems which are
established for Diophantine sets $\D(\phi)$ may also be established
(usually in slightly weaker form) for the corresponding larger, nicer
sets $\D(\phi,R)$. The standard technique for doing so involves
removing the ``$R$-tail" of a series before applying Diophantine
conditions, then checking that the tail is small.  This technique was
called the ``ultraviolet cutoff" by Arnold in his proof of the KAM
theorem [Ar], and is illustrated in the proof of Lemma 2 below.

%%%\vfill\eject %%% pagebreak
\medskip
\noi{\bf 4.2.2 \ Besjes' inequality generalized to functions far from
low-order resonance}

\proclaim Lemma 2. Let $S=\Rn\times\R$, and
suppose $g:S\to\Rn$ satisfies assumptions (j), (jj) from
Subsection 2.2, along with assumption (jw) from Subsection 2.3.  Let the
zone function $\phi$ be adapted to $g$ on $\Rn$ in the sense of
Eq.\ (4.2), and define the positive constants $C_1=C_1(g,\phi)$ and
$C_2=C_2(g,\phi)$ by
$C_1 =2\sum_{0\not=k}{\|g_k\|_\Rn/\phi(|k|)}$ and
$C_2 =\sum_{0\not=k}{\|Dg_k\|_\Rn/\phi(|k|)}$.
Let $\nu\in\D(\phi,R)$.  If
$\{x_n\}_{n=0}^\infty\subset \Rn$ is a sequence for which the successive
differences $x_{n+1}-x_n$ are bounded by $M$ (i.e.,
$\sup_n|x_{n+1}-x_n|\le M$), then
$$
 \biggl|\sum_{n=0}^{N-1} g(x_n,n\nu)\biggr|
 \ \le \
 C_1
 \,+\,
 N\Bigl(C_2 M  + \sum_{|k|>R}\|g_k\|_\Rn\Bigr)\,,
 \qquad {\rm where} \qquad
 \sum_{|k|>R}\|g_k\|_\Rn\to0 \quad {\rm as} \quad R\to\infty.
$$

\noi{\it Proof.}
Since $C_1<\infty$, we write $g$ as its uniformly convergent Fourier
series
$ g(x,\theta)  = \sum_{0\not=k\in\Z} g_k(x) e^{2\pi ik\theta}$,
so that
$$
\biggl|\sum_{n=0}^{N-1} g(x_n,n\nu)\biggr|
 \ \le \
 \biggl|\sum_{n=0}^{N-1}\,\sum_{0<|k|\le R}
 g_k(x_n) e^{2\pi i kn\nu}
 \biggr|
 \ + \
 \biggl|\sum_{n=0}^{N-1}\,\sum_{|k|>R}
 g_k(x_n) \,e^{2\pi i kn\nu}
 \biggr|\,.
\eqno(4.4)
$$\vn
We shall treat separately each of the double sums on the right-hand
side of inequality (4.4).  For the first double sum, we reverse the order of
summation and use the ``summation by parts" formula \hfill\break
$\sum_{n=0}^{N-1}a_n(b_{n+1}-b_n) =
(a_N b_N - a_0 b_0) - \sum_{n=0}^{N-1}(a_{n+1} - a_n)b_{n+1}$
with $a_n = g_k(x_n)$ and $b_n = e^{2\pi ikn\nu}/(e^{2\pi ik\nu}-1)$
so that $a_n(b_{n+1}-b_n) = g_k(x_n) e^{2\pi i kn\nu}$.
It then follows that
$$
 \biggl|\sum_{0<|k|\le R}\,\sum_{n=0}^{N-1}
 g_k(x_n) e^{2\pi i kn\nu}
 \biggr|
 \ \le \
 \sum_{0<|k|\le R}
\,\biggl|\,
{g_k(x_N) e^{2\pi iNk\nu} - g_k(x_0)
\over
e^{2\pi ik\nu}-1}
 \ \ - \
\sum_{n=0}^{N-1}
\bigl(g_k(x_{n+1})-g_k(x_n)\bigr)
{ e^{2\pi i(n+1)k\nu}
\over
e^{2\pi ik\nu}-1}
\,\biggr|
$$\vn
$$
\le \
\sum_{0<|k|\le R}
\,\biggl(
{2\|g_k\|_\Rn
\over
|e^{2\pi ik\nu}-1|}
 \ + \,
{\|Dg_k\|_\Rn
\over
|e^{2\pi ik\nu}-1|}
\sum_{n=0}^{N-1}
|x_{n+1}-x_n|
\,\biggr)
 \ \le \
\sum_{0<|k|\le R}
{2\|g_k\|_\Rn
+
NM\|Dg_k\|_\Rn
\over
|e^{2\pi ik\nu}-1|}
$$\vn
$$
 \le \
 \sum_{0<|k|\le R}
 {2\|g_k\|_\Rn
 +
 NM\|Dg_k\|_\Rn
 \over
 \phi(|k|)}
 \ \, \le \ \
 \sum_{0\not=k}
 {2\|g_k\|_\Rn
 +
 NM\|Dg_k\|_\Rn
 \over
 \phi(|k|)}
 \ = \
 C_1
 \,+\,
 NMC_2 \,.
 \eqno(4.5)
$$\vn
We next treat the second double sum (the $R$-tail) on the right-hand
side of inequality (4.4) using the simple estimate
$$
 \biggl|\sum_{n=0}^{N-1}\,\sum_{|k|>R}
 g_k(x_n) \,e^{2\pi i kn\nu}
 \biggr|
 \ \le \
 N\sum_{|k|>R}\|g_k\|_\Rn
 \ \to \ 0
 \quad {\rm as} \quad
 \ R \ \to \ \infty\,.
 \eqno(4.6)
$$\vn
Inserting estimates (4.5) and (4.6) into inequality (4.4) concludes
the proof.\quad//

\smallskip
\noi{\bf Remark 4.6} \ A related analogous result for flows (but
without the ultraviolet cutoff) appears as Lemma 13 of [S\'a],
and in Theorem 2 of [ES], and a more general Besjes-type inequality for
so-called KBM vector fields also appears in [S\'a] as Lemma 2.
A still more closely related result for flows appears as
Lemma 2 in our previous paper [DEG], where it was used in averaging
methods applied to certain classes of charged particle motions in
crystals.

\smallskip
\noi{\bf Remark 4.7} \ In the case where $g$ has a finite Fourier series,
the above proof simplifies in obvious ways; but these simplifications
become problematic as the Fourier series grows in length (note that the
example in \S 3.3 has a Fourier series with only four terms).

\vfill\eject %%% pagebreak

\medskip\noi
{\bf 4.2.3 Proof of Theorem 2}

\smallskip
Assume the hypotheses of Theorem 2 (cf.\ \S 2.2; note that assumption (j)
ensures the existence of zone functions adapted to $f$, as discussed before
Remark 4.4). The proof is essentially the same as the proof of Theorem 1 with
appropriate changes as needed in order to use Lemma 2.  As in the previous
proof, the solutions $x_N$ and $y_N$ clearly exist uniquely for all $N\in\N$.
For the approximation relation, we write as before
$$
 |x_N-y_N|
  \le \
 \eps L \sum_{n=0}^{N-1}|x_n-y_n|
 \ + \
 \eps \,
  \Bigl|
 \sum_{n=0}^{N-1}
 \fwig(y_n,n\nu)
 \Bigr|
$$\vn
where $\fwig(y,\theta):= f(y,\theta) - \fbar(y)$ is the oscillating part of
$f$.  The hypotheses clearly imply that $\|f\|_S<\infty$, and since $\phi$ is
adapted to $f$ on $\Rn$, the constants $C_1$ and $C_2$ from Lemma 2 are well
defined. We may thus set $C=(C_1+C_2T\|\fbar\|_\Rn+1)e^{LT}$.  Finally, we fix
the parameter $\CT>0$ and choose $R_\eps>0$ so large
that $\sum_{|k|>R_\eps}\|f_k\|_\Rn \le \CT\eps$, where $f_k(x)$ is the $k$th
Fourier coefficient of $f$.  It is now a simple matter to check that if
$\nu\in\D(\phi,R_\eps)$, then the hypotheses of Lemma 2 are satisfied with
$M:=\eps\|\fbar\|_S$.  We thus have
$$
 |x_N-y_N|
 \ \le \
 \eps\, L \sum_{n=0}^{N-1}|x_n-y_n|
 \ + \
 \eps C_1
 \,+\,
 \eps N\Bigl(C_2 M  + \sum_{|k|>R}\|f_k\|_\Rn\Bigr)
$$\vn
$$
 \ \le \
 \eps\, L \sum_{n=0}^{N-1}|x_n-y_n|
 \ + \
 \eps C_1
 \,+\,
 \eps N\bigl(C_2 \eps\|\fbar\|_\Rn  + \CT \eps\bigr)\,,
$$\vn
and so for $0<N\le T/\eps$, we have
$\disp
|x_N-y_N|
 \ \le \
 \eps L \sum_{n=0}^{N-1}|x_n-y_n|
 \ + \
 \eps\bigl(C_1 + C_2 T \|\fbar\|_\Rn  + \CT\bigr)
$.
Applying Gronwall's inequality for sequences (Lemma 3,
Appendix) gives
$
|x_N-y_N|
\ \le \
\eps \bigl(C_1 + C_2 T \|\fbar\|_\Rn  + \CT\bigr) e^{\eps L N}
\ \le \
\eps \bigl(C_1 + C_2 T \|\fbar\|_\Rn  + \CT\bigr) e^{L T}
\ = \
C \eps
$
for
$0<N\le T/\eps$, as claimed.
The second part of Theorem 2 (namely $|x_n-y(n)|\le C'\eps$ for
$0\le n\le T/\eps$) again follows from Lemma 4 (Appendix) and
the triangle inequality.\quad//

\smallskip
\noi{\bf Remark 4.8} \ It is important to note that for fixed positive
$\CT$ and $\eps$, the
ultraviolet cutoff $R_\eps$ need not be very large to ensure that
$\sum_{|k|>R_\eps}\|f_k\|_\Rn\le \CT\,\eps$, whence the number of
inequalities to be checked in Eq.\ (4.3) (with $R=R_\eps$) is also modest.
In fact, straightforward estimation shows that when the Fourier
coefficients of $f$ decrease as
$\|f_k\|_\Rn \le C |k|^{-(p+1)}$ \ (e.g. when $f$ is of class $C^{p+1}$),
it is enough to take
 $R_\eps \ge 1 + \bigl({2C\over p\CT\eps}\bigr)^{1/p}$
 (and when the coefficients decrease as
$\|f_k\|_\Rn \le C e^{-\rho|k|}$, it is enough to take
$R_\eps \ge 1 + \ln\bigl({2C\over\rho \CT\eps}\bigr)^{1/\rho}$).

\medskip
\noi{\bf Remark 4.9} \ If an $O(\eps^2)$ term is added to Eq.\ (1.4)
 so that it reads \
$
x_{n+1} \ = \ x_n + \eps f(x_n,n\nu) + \eps^2 g(x_n,n\nu)\,,
$
where $g:S\to\Rn$ satisfies the hypotheses of Theorem 2,
then it is a simple matter to check that Theorem 2 continues to
hold with the order constant $C$ replaced by
$C'= (C_1 + C_2T\|\fbar\|_\Rn + \CT + \|g\|_S)e^{LT}$.  This form of
Theorem 2 is often useful in applications.

\medskip\noi
{\bf 4.2.4 Proof of Theorem 3}

\smallskip
Assume the hypotheses of Theorem 3 (these include those of Theorem 2
together with the additional zero-mean assumption (jw); cf.\
\S 2.3).  The hypotheses clearly imply that $\|f\|_S<\infty$, and since
$\phi$ is adapted to $f$ on $\Rn$, the constants $C_1$ and $C_2$ from the
conclusion of Lemma 2 are well defined. We may thus
choose the parameter $\CT>0$ and set $K_1=C_1$ and
$K_2 = C_2\|f\|_S+\CT$.  Finally, we choose $R_\eps>0$ so large that
$\sum_{|k|>R_\eps}\|f_k\|_\Rn \le \CT\,\eps$.  It is now a simple matter
to check that whenever $\nu\in\D(\phi,R_\eps)$, the hypotheses of
Lemma 2 are satisfied with $M:=\eps\|f\|_S$, from which we conclude
that
$$
 |x_N - x_0| \ = \
 \eps\,\biggl|
 \sum_{n=0}^{N-1}
 f(x_n,n\nu)
 \biggr|
 \ \le \
 \eps C_1 + \eps N\Bigl(C_2M + \sum_{|k|>R_\eps}\|f_k\|_\Rn\Bigr)
$$\vn
$$
 \le \
 \eps\,C_1 + \eps N\bigl(C_2\eps\|f\|_S + \CT\eps \bigr)
 \ \le \
 K_1\eps + K_2\eps^2 N.\quad//
$$

\smallskip
\noi{\bf Remark 4.10} \ The proof of Theorem 3 is so short, and
its hypotheses are so closely related to those of Lemma 2, that
it is nearly a corollary of Lemma 2.  The interesting features of
Theorem 3 are that long-time invariance is
shown without the traditional transformation of variables, while
$\nu$ is required to be Diophantine only to low order $R_\eps$.

\bigskip
\noi{\bf 4.3 Proofs of Propositions A, B, and C}

\smallskip
For the statements of Propositions A, B, and C, see Subsection 2.4.

\medskip
\noi{\bf 4.3.1 Proof of Proposition A}

\smallskip
The zone functions enter the proofs of Theorems 2 and 3 only through
Lemma 2.  It is clear that if $\phi$ is replaced by $\eps^\lambda\phi$ in
Eq.\ (4.5), then the final estimate of Lemma 2 is changed to
$\eps^{-\lambda}(C_1+NMC_2)$.
The error bound in Theorem 2 then changes to
 $|x_N-y_N|\le\eps^{1-\lambda}(C_1+C_2T\|\fbar\|_\Rn)e^{LT}
  + \eps \CT e^{LT} =O(\eps^{1-\lambda})$ for $0\le N\le T/\eps$,
while the error bound in Theorem 3 changes to
 $|x_N-x_0| \le \eps^{1-\lambda} C_1
  + \eps N\bigl(C_2\eps^{1-\lambda}\|f\|_S + \CT\eps \bigr)
  \le K_1\eps^{1-\lambda} + K_2\eps^{2-\lambda} N$.\quad//

\medskip
\noi{\bf 4.3.2 Proof of Proposition B}

\smallskip
Here we give the proof of Proposition B as it applies to Theorem 1 only;
the proofs of its applicability to Theorems 2 and 3 are nearly the same.

Fix $\eps>0$, let $U\subset\Rn$ be open, take $S'=U\times\N$, and suppose
$g:S'\to\Rn$, where $g$ is not assumed to have compact support in $U$ (in
other words, $g$ satisfies assumptions (i) and (ii) of \S 2.1, with $S'$
in place of $S$, but does not satisfy assumption (iii)).

We now use $g$ to define the systems $(1')$, $(2')$, and $(3')$,
which are simply the previous systems (1.1), (1.2), and (1.3),
respectively, in which the perturbation $\eps f$ has been replaced
by $\eps g$.  We assume that the common initial condition
$x_0=y_0=y(0)$ is fixed in $U$, and we choose the positive
timescale parameter $T<\beta(x_0)$,  where $[0,\beta(x_0))$ is the
maximal forward interval of existence for the initial value problem
$$
{d\yhat\over dt'} \, = \,
{\widehat{g}}(\yhat)\,,
\qquad
\yhat(0) = x_0\in U,
\eqno(3'')
$$\vn
which is simply the scaled, $\eps$-independent version of system $(3')$
obtained by introducing the ``slow time" $t'=\eps t$.
We then let $Z = \{z\in U \,|\, z=\yhat(t'), \ 0\le t'\le T\}$ denote the
solution curve of system $(3'')$ over $[0,T]$, and we choose $\delta>0$
such that $\delta<{\rm dist}(Z,\partial U)$.
Then the closure ${\overline D}(\delta)$ of the open ``$\delta$-tube''
$D(\delta)$
around $Z$ formed by the union of open balls of radius $\delta$
having centers in $Z$ is contained in $U$;  i.e.,
$D(\delta) := \bigcup_{t\in[0,T]} B_\delta\bigl(y(t)\bigr)
  \ \subset \ {\overline D}(\delta) \ \subset \ U,$
where $B_\delta(y)$ denotes the open ball of radius $\delta$ centered on
$y$ in $\Rn$.

We next choose $r>0$ so that the open ball $B_r(x_0)$ contains
${\overline D}(\delta)$, and we define the
compactly supported function $f:\Rn\times\N\to\Rn$ which
 \ (a) coincides with $g$ on ${\overline D}(\delta)\times\N$,
 \ (b) vanishes on $B_r(x_0)^{\rm c}\times\N$ (here $^{\rm c}$
     denotes ``complement"), and
 \ (c) interpolates $g$ on $B_r(x_0)\cap{\overline D}(\delta)^{\rm c}$
 in such a way that $f$ is of the same smoothness class as $g$ and such that
$\|f\|_{\Rn\times\N}=\|g\|_{{\overline D}(\delta)\times\N}$.  The existence
of such $f$ is guaranteed by the ``smooth Tietze extension theorem" as
given, for example, on p.\ 380 of [AMR].

Using this $f$, and the constant $T$ (from the existence interval $0\le
t'\le T$ of the solution $\yhat(t')$ of $(3'')$, corresponding to the
existence interval $0\le t\le T/\eps$ for the solution $y(t)=\yhat(\eps
t)$ of $(3')$), we apply Theorem 1 and Lemma 4 from the Appendix to
conclude that, for appropriate $C_1, C_2 >0$, we have:
$$
|x_n - y_n| \le C_1\,\eps
\qquad {\rm for} \qquad
0\le n\le T/\eps\,,
\quad
y_n\in D(\delta/2)\,,
\quad {\rm and} \quad
x_n\in D(\delta)\,;
\qquad {\rm and}
$$\vn
$$
|y_n-\yhat(\eps n)| \le C_2\,\eps
\qquad {\rm for} \qquad
0\le n\le T/\eps
\quad {\rm and} \quad
y_n\in D(\delta/2)\,.
$$\vn
Using these inequalities together with the triangle inequality, if we now
impose a smallness condition on $\eps$ by requiring it to be strictly
less than the threshold $\eps_0 := {\rm min}\{\delta/(2C_1),
\delta/(2C_2)\}$, we find that the conditions $y_n\in D(\delta/2)$ and
$x_n\in D(\delta)$ are ensured for $0\le n\le T/\eps$, and it follows that
$|x_n-\yhat(\eps n)|\le(C_1+C_2)\eps<\delta$ also holds for $0\le n\le
T/\eps$. Finally, since $x_n$, $y_n$, and $y(n)=\yhat(\eps n)$ remain in
$D(\delta)$ for $0\le n\le T/\eps$, and since $f$ and $g$ coincide on
$D(\delta)$, we see that whenever $0\le\eps<\eps_0$, the dynamics of
systems $(1')$, $(2')$, and $(3')$ coincide with the dynamics of the
respective systems (1.1), (1.2), and (1.3) on the interval $0\le n\le
T/\eps$, which completes the proof.\quad//

\smallskip
\noi{\bf Remark 4.11} \ In the above proof, the order constants $C_1$,
$C_2$ and the threshold $\eps_0$ depend on $\delta$.
We note that, since the motions of systems (1.1), (1.2), and (1.3)
remain in the $\delta$-tube $D(\delta)$, the uniform
norms which appear in the proofs of Theorem 1, 2, and 3 may be taken over
${\overline D}(\delta)$ rather than all of $\Rn$.

\vfill\eject %%% pagebreak

\medskip
\noi{\bf 4.3.3 Proof of Proposition C}

\smallskip
Let $g(u,n):=(f(x,nq/p+\tau),\,a)\transp$ where $u:=(x,\tau)\transp$,
so that $g : U\times\R\times\N\to\R^{d+1}$.
The system $u_{n+1} = u_n +\eps g(u_n,n)$ clearly satisfies the hypotheses of
Proposition B applied to Theorem 1, with $d$ replaced by $d+1$ and $U$
replaced by $U\times\R$.
Thus the conclusion of Theorem 1 applies to $u_n$ as well as to $x_n$.
The constants $c$ and $c'$ may be easily estimated along the lines of the
proofs of Theorem 1 and Lemma 4 respectively.
Taking into account Remark 4.11, we find
$ c(T,|a|)=\bigl( L_g T \|{\widehat g}\|_{\overline D}
     +
 \|{\widetilde g}\|_{\overline{D}\times\N} \bigr) e^{L_g T}$
 \ and \
$c'(T,|a|)=TL_g \|{\widehat g}\|_{\overline D}\,e^{L_gT}$.
Here $L_g$ is the $u$-Lipschitz constant of $g$, which is independent of
$a$.  On the other hand, the norms $\|{\widehat g}\|_{\overline D} $ and
$\|{\widetilde g}\|_{\overline{D}\times\N}$ depend on $|a|$, since
$\|{\widehat g}\|_{\overline D} = \sup_{v\in {\overline D}}
 \sqrt{|{\widehat f}(v)|^2+a^2}$
 \ and \
$\|{\widetilde g}\|_{\overline{D}\times\N} = \sup_{v\in {\overline D},
 n\in\N} \sqrt{|{\widehat f}(v)-f(x,nq/p+\tau)|^2+a^2}$.\quad//

\bigskip\medskip\noi
{\bf Appendix.}

\medskip\noi
In this appendix, for the sake of completeness we supply statements
and proofs of two elementary results with which the reader may be unfamiliar.

\smallskip
\proclaim Lemma 3 (The Gronwall inequality for sequences).
Let $A\ge0$, $B\ge0$, and $\{E_n\}_{n=0}^\infty$ be a sequence of
nonnegative real numbers with $E_0=0$  satisfying
$\displaystyle E_N \ \le \ A \sum_{n=0}^{N-1}E_n \ + \ B$.
Then $E_N \,\le\, Be^{AN}$.

\noi{\it Proof.} \ Set $R_{N-1} =A \sum_{n=0}^{N-1}E_n \ + \ B$
so that $R_N - R_{N-1} = A E_N \le A R_{N-1}
\Rightarrow R_N \le (1+A)R_{N-1}$.  Proceeding inductively,
we find that
$R_N \le (1+ A)R_{N-1}\le\ldots\le (1+A)^N R_0 =
B(1+A)^N \le B e^{AN}$,
where we have used $R_0 = B$ and
$x>0\Rightarrow (1+x)^{1/x}\le e$.\quad//

\smallskip
\proclaim Lemma 4 (Equivalence of autonomous flows and maps).  Let
$\eps>0$, and suppose $\fbar:\Rn\to\Rn$ is Lipschitz continuous and
has compact support.  Then
$$
{\sl the \ map} \qquad
y_{n+1} = y_n + \eps\fbar(y_n) \qquad (1.5)
\qquad\qquad {\sl and \ the \ flow} \qquad
{dy\over dt} = \eps\fbar(y) \qquad (1.6)
$$\vn
are equivalent in the sense that there exists a constant
$K>0$ such that the solutions $y_n$ and $y(t)$ of (1.5) and (1.6),
respectively, with common initial condition $y_0 = y(0) \in\Rn$
satisfy the nearness condition
$|y_n - y(n)| \;\le\; K\eps$ \ for \ $0\le n\le T/\eps$.

\medskip
\noi{\it Proof.} \ Let $L>0$ denote the global Lipschitz constant of
$\fbar$. First we note that $y(n+1) - y(n) = \eps \int_n^{n+1}
\fbar(y(t))\,dt=
 \eps\fbar(y(n)) +
\eps\int_n^{n+1} \bigl(\fbar(y(t))-\fbar(y(n))\bigr) \,dt$.
Thus $y_{n+1} - y(n+1) =
 y_n - y(n) + \eps\bigl(\fbar(y_n) - \fbar(y(n))\bigl)
 -\eps\int_n^{n+1} \bigl(\fbar(y(t))-\fbar(y(n))\bigr) \,dt$.
Now setting $E_n = |y_n - y(n)|$, we obtain
$E_{n+1} \le E_n + \eps L E_n + \eps^2 L \|\fbar\|_{\Rn}$, since
$\eps|\int_n^{n+1} \bigl(\fbar(y(t))-\fbar(y(n))\bigr) \,dt| \le
 \eps L \int_n^{n+1} |y(t)-y(n)| \,dt \le \eps^2 L \|\fbar\|_{\Rn}$.
Using this last inequality to form a telescoping sum, we arrive to
 $E_n - E_0 \le \eps L\sum_{k=0}^{n-1} E_k + n\eps^2 L \|\fbar\|_{\Rn}$,
\ or \ $E_n \le \eps L\sum_{k=0}^{n-1} E_k + \eps TL \|\fbar\|_{\Rn}$
 (since $E_0=0$ and $0\le n\le T/\eps$).
Finally, we apply the Gronwall inequality for sequences (Lemma 3,
above) to get $E_n \le \eps TL \|\fbar\|_{\Rn}\,e^{\eps Ln} \le
  \eps TL \|\fbar\|_{\Rn}\,e^{LT}$, so the
desired conclusion is true with $K = TL \|\fbar\|_{\Rn}\,e^{LT}$.\quad//

\bigskip\medskip
\noi{\bf Acknowledgments.} HSD thanks F.\ Golse and L.\ Michelotti
for stimulating discussions, and acknowledges the support of the
C.P.\ Taft Foundation at the University of Cincinnati.
JAE and MV gratefully acknowledge discussions with
L. Michelotti and T. Sen, as well as
support from DOE grant DE-FG03-99ER41104.

\bigskip\medskip
\noi{\bf References}

\parindent = 40 true pt
\parskip = 6 true pt

\item{[AMR]}R. Abraham, J. Marsden, and T. Ratiu, {\it Manifolds, Tensor
Analysis, and Applications} (2nd Ed.), Springer-Verlag, New York, 1988.

\item{[ABG]}M. Andreolli, D. Bambusi, and A. Giorgilli, On a weakened
form of the averaging principle in multifrequency systems, {\it
Nonlinearity} {\bf 8} (2): 283--293 (1995).

\item{[Ar]} V.I. Arnold, Proof of A.N. Kolmogorov's theorem on the
preservation of quasi-periodic motions under small perturbations of
the Hamiltonian [Russian], {\it Usp.\ Mat.\ Nauk.\ SSSR} {\bf
18} (5): 13--40 (1963) [English translation: {\it Russian Math.\
Surveys} {\bf 18} (5): 9--36 (1963)].

%\item{[BMT]}  A. Bazzani, S. Marmi, G. Turchetti, Nekhoroshev estimates for
%nonresonant symplectic maps, {\it Celestial Mechanics} {\bf 47}, 333
%(1990).

\item{[BGSTT]}  A. Bazzani, M. Giovannozzi, G. Servizi, G. Turchetti,
 E. Todesco, Resonant normal forms and stability analysis for area
 preserving maps, {\it Physica D} {\bf 64}, 66 (1993).

%\item{[BTT]}A. Bazzani, E. Todesco, and G. Turchetti, A normal form approach
%to the theory of nonlinear betatronic motion, {\it CERN Report} 94--02,
%Geneva, 1994.

\item{[Bel]}E.P. Belan,
On the averaging method in the theory of finite difference equations
[Russian], {\it Ukrain. Mat.\ Z.}\ {\bf 19}, no.\ 3: 85--90 (1967).

\item{[Bes]}J. Besjes, On the asymptotic methods for non-linear
differential equations, {\it J.\ M\'ecanique} {\bf 8}:
357--372 (1969).

\item{[BM]}N.N. Bogoliubov and Y.A. Mitropolsky, {\it  Asymptotic
Methods in the Theory of Non-Linear Oscillations} (2nd Ed.)
[translated from Russian], Gordon and Breach Science Publishers,
New York, 1961.

\item{[BHS]}H.W. Broer, G.B. Huitema, and M.B. Sevryuk, {\it
Quasiperiodic Motions in Families of Dynamical Systems}, Lecture
Notes in Mathematics, Vol.\ 1645, Springer-Verlag, New York, 1996.

\item{[Br]}A. Browder, {\it Mathematical Analysis}, Springer-Verlag,
New York, 1996.

\item{[CBW]}A.W. Chao, P. Bambade and W.T. Weng, Nonlinear Beam-Beam
 Resonances, in {\it Lecture Notes in Physics} {\bf
247}, 77--103, Springer-Verlag, New York, 1986.

\item{[Dr]}V.A. Dragan,
Method of averaging for systems of sum-difference equations [Russian],
{\it Mat.\ Issled.} ({\it Computational Methods of Mechanics}), No.\ 64:
172--181, 195--196 (1981);
Methods of averaging and freezing of systems of finite difference
equations of two variables [Russian], {\it Mat.\ Issled.}, No.\ 64:
182--188, 196--197 (1981).

\item{[DEG]}H.S. Dumas, J.A. Ellison, and F. Golse, A mathematical
theory of planar particle channeling in crystals,
{\it Physica D\/} {\bf 146} (1--4): 341--366 (2000).

\item{[DESV]}H.S. Dumas, J.A. Ellison, T. Sen, and M. Vogt, work in
preparation.

\item{[DEV]}H.S. Dumas, J.A. Ellison, and M. Vogt, presentation at
2002 Spring Meeting of APS, in Albuquerque, NM; see \
{\tt http://www.math.unm.edu/{\~\ $\!\!\!$}ellison/papers/APS02\_MAP.ps.gz}

\item{[ES]}J.A. Ellison and H.-J. Shi, The method of averaging
in beam dynamics, in {\it Accelerator Physics Lectures at the
Superconducting Super Collider} (AIP Conf.\ Procs.\ 326,
Y. Yan and M. Syphers, Eds.): 590--632 (1995).

\item{[Fo]}\'{E}. Forest, {\it Beam Dynamics: A New Attitude and Framework},
 Harwood Academic Publishers, Amsterdam, 1998.

\item{[K\"o]}T.W. K\"orner, {\it Fourier Analysis}, Cambridge University
Press, Cambridge, 1988.

% \item{[Ne1]}A. Neishtadt, On the accuracy of persistence of adiabatic
% invariant in single-frequency systems,
% {\it Regul.\ Chaotic Dyn.} {\bf 5}, no.\ 2: 213--218 (2000);
% The separation of motions in systems with rapidly rotating phase,
% {\it J. Appl.\ Math.\ Mech.}\ {\bf 48} (1984), no. 2: 133--139 (1985),
% translated from Russian {\it Prikl.\ Mat.\ Mekh.} {\bf 48}, no.\ 2:
% 197--204 (1984).

\item{[Ne]}A. Neishtadt (private communication w/HSD), June, 2000.

\item{[PW]}R.E.A.C. Paley and N. Wiener, {\it Fourier Transforms in
the Complex Domain}, AMS Colloquium Publications, Vol.\ 19, New York,
1934.

\item{[Ru]}R.D. Ruth, Single Particle Dynamics and Nonlinear Resonances
 in Circular Accelerators, in {\it Lecture Notes in Physics} {\bf
247}, 37--63, Springer-Verlag, New York, 1986.

\item{[R\"u]}H. R\"u{\ss}mann, On optimal estimates for the solutions
of linear partial differential equations of first order with constant
coefficients on the torus, in {\it Lecture Notes in Physics} {\bf
38}, 598--624, Springer-Verlag, New York, 1975;
On the frequencies of quasi periodic solutions of analytic integrable
Hamiltonian systems, in {\it Seminar on Dynamical Systems}
(Proceedings of the Euler International Mathematical Institute, St.\
Petersburg, 1991; V. Lazutkin et al., Eds.), 160--183, Birkh\"auser,
Berlin, 1994.

\item{[S\'a]}A.W. S\'aenz, Higher order averaging for nonperiodic
systems, {\it J.\ Math.\ Phys.}\ {\bf 41}: 5342--5368 (2000).

% \item{[SV]}J.A. Sanders and F. Verhulst, {\it Averaging Methods in
% Nonlinear Dynamical Systems}, Applied Mathematical Sciences, Vol.\ 59,
% Springer-Verlag, New York, 1985.

\item{[Yo]}J.-C. Yoccoz, An introduction to small divisors problems, in
{\it From Number Theory to Physics} (Les Houches, 1989), 659--679,
Springer-Verlag, Berlin, 1992.

\bye